\documentclass[pra, a4paper, amsfonts, amssymb, amsmath, reprint, showkeys, nofootinbib, oneside, superscriptaddress,floatfix,aps]{revtex4-2}
\usepackage[english]{babel}
\usepackage[utf8]{inputenc}
\usepackage[colorinlistoftodos, color=green!40, prependcaption]{todonotes}
\usepackage{amsthm}
\usepackage{mathtools}
\usepackage{placeins}
\usepackage{physics}
\usepackage[dvipsnames]{xcolor}
\usepackage{graphicx}
\usepackage[left=18mm,right=18mm,top=35mm,columnsep=15pt]{geometry}
\usepackage{adjustbox}
\usepackage{placeins}
\usepackage[T1]{fontenc}
\usepackage{lipsum}
\usepackage{csquotes}
\usepackage{siunitx}
\usepackage{caption}
\usepackage{multirow}
\usepackage{subcaption}
\usepackage[pdftex, pdftitle={Article}, pdfauthor={Author}, colorlinks=true]{hyperref}
\usepackage[normalem]{ulem}

\begin{document}
\title{ 
Spatial superposition for a two-dimensional matter-wave interferometer in an inverted harmonic potential with gyroscopic rotational stability
}

\author{Ryan Rizaldy}
    \affiliation{Van Swinderen Institute for Particle Physics and Gravity, University of Groningen, 9747AG Groningen, the Netherlands}

\author{Tian Zhou}
    \affiliation{Van Swinderen Institute for Particle Physics and Gravity, University of Groningen, 9747AG Groningen, the Netherlands}

\author{Run Zhou}
    \affiliation{Key Laboratory of Low-Dimensional Quantum Structures and Quantum Control of Ministry of Education, Key Laboratory for Matter Microstructure and Function of Hunan Province, Department of Physics and Synergetic Innovation Center for Quantum Effects and Applications, Hunan Normal University, Changsha 410081, China}

\author{Anupam Mazumdar}
    \affiliation{Van Swinderen Institute for Particle Physics and Gravity, University of Groningen, 9747AG Groningen, the Netherlands}


\begin{abstract}
This study presents a mathematical model of the spatial and rotational motion of a nanodiamond in an inverted harmonic potential to create a macroscopic quantum spatial superposition. The model is based on the Stern-Gerlach Interferometer (SGI) scheme, which utilises linear and quadratic magnetic fields to generate a harmonic potential (linear magnetic field) and a non-linear potential (non-linear/quadratic magnetic field). By incorporating two-dimensional dynamics into the model, we provide a more realistic and accurate depiction of nanoparticle dynamics in linear and inverted harmonic potentials and explore the interaction between motion in a two-dimensional plane.  Importantly, we derive the equations of motion for the rotational degrees of freedom, i.e. libration, precession, and rotation. The results show that adding a magnetic-field bias term to the magnetic-field profile in the linear stage affects the classical equations of motion but does not affect the width of the wave packet. Moreover, the libration mode always forms a harmonic potential at each stage because the applied initial angular velocity is dominated by the nanoparticle's defect axis, making it more stable in the presence of the trap frequency in the orthogonal direction along the axis that enables the creation of a macroscopic quantum superposition.

\end{abstract}

\maketitle

\section{INTRODUCTION}


The experimental realisation of a macroscopic quantum spatial superposition has become a cornerstone in exploring the boundaries between quantum and classical physics. Generating such spatial superpositions is pivotal for testing the validity of quantum mechanics at larger scales and for investigating fundamental phenomena such as wave-function collapse and gravitational decoherence~\cite{AdlerBassi2009,
bassi2013collapse, Pfister:2015sna, Bassi:2017szd, Anastopoulos:2021jdz,Xu:2020lhc,Toros:2020krn, Petruzziello:2020wkd}. Experimental progress in fields such as optomechanical systems has enabled the extension of quantum effects 
from electrons, atoms to macromolecules
to macroscopic length scales \cite{TonomuraEtAl1989,Kasevich:1989zz,
Kovachi_2015,asenbaum2017phase,Amit2019, Folman2018,doi:10.1126/sciadv.abg2879,
Arndt_1999,arndt, arndt2014superpositions}. These precision led experiments, coupled with precise control of environmental decoherence~\cite{romeroisart2011collapse,Romero-Isart:2011yun}, have laid the groundwork for the realization of quantum states of increasingly massive objects \cite{scala2013matter,Pedernales:2020nmf,Wan16_GM,steiner2024pentacene,Marshman2022,Zhou:2022epb,Zhou:2022frl,Zhou:2024voj,Zhou:2022jug,Kialka2022Roadmap}. One of the pressing fundamental application to create macroscopic quantum superposition is to test the quantum nature of gravity in a lab via witnessing the spin entanglement~\cite{ICTS,Bose:2017nin}, see also~\cite{Marletto:2017kzi}. The protocol is known as the QGEM (quantum gravity induced entanglement of matter)~\cite{Marshman:2019sne,Bose:2022uxe,Vinckers:2023grv}. Motivated by the QGEM protocol, one can also experimentally probe the quantum analogue of the light-bending test due to gravity, thereby witnessing entanglement between matter and photon via the exchange of a virtual massless graviton~\cite{Biswas:2022qto}. One can also test the simplest models of massive gravitons~\cite{Elahi:2024dbb}, and the quantum version of the equivalence principle~\cite{Bose:2022czr}.

To test the QGEM protocol, we require mass 
of order $m\sim 10^{-15}-10^{-14}$~kg and spatial superposition of order $\Delta x \sim 1-50~{\rm \mu m}$, see~\cite{Bose:2017nin,Schut:2023hsy,Schut:2025blz}. To realize these experiments, one challenge is to create a large spatial superposition. This paper will aim to study the creation of large spatial superposition by employing an inverted harmonic potential (IHP) as the “inflation” stage in a Stern–Gerlach interferometer (SGI), motivated by our earlier paper~\cite{Zhou:2024voj}. The initial motivation comes from the paper ~\cite{RomeroIsart2017_CoherentInflation, PhysRevLett.132.023601,pino2018chip}, which proposed the initial idea of creating a large spatial superposition.
The advantage of IHP is the realization of a large spatial superposition due to tachyonic instability, see~\cite{Zhou:2024voj,Moorthy:2025bpz}. Furthermore, 
to realize the SGI, we also need to levitate the neutral nanoparticle diamagnetically on a chip, see~\cite{Schut:2023hsy,
Elahi:2024dbb} by following the initial experimental realization of levitating the nanoparticle by a fixed magnet~\cite{DUrso16_GM}.

One of the complications that arises in creating a spatial superposition of macroscopic objects is that of the rotational degrees of freedom along with the motional ones. For instance, in the case of a single NV (nitrogen vacancy) centered nanodiamond in the presence of an external magnetic field, the nanodiamond experiences torque, NV-spin libration, and the well-known Einstein–de Haas effect; see~\cite{stickler2021quantum}, see ~\cite{Japha:2022phw, japha2021unified}. Consequently, the achievable superposition size is also influenced by the specific spin configurations in the two interferometer arms, i.e., left and right arms of the matter-wave interferometer. Rotational phenomena are known to play a crucial role in nanocrystals; see the key references~\cite{chen2019nonadiabatic,Stickler18_GM,Rusconi:2022jhm,stickler2021quantum,ma2021torque,jin2024quantum,Kuhn2017,wachter2025gyroscopicallystabilizedquantumspin}, and a review on rotational aspects of a nanoparticle~\cite{Rademacher:2025sye}.

In addition, rotational effects have been shown to alter both the spatial extent of the superposition and the observed spin contrast~\cite{Japha:2022phw}, a situation often referred to as the Humpty-Dumpty problem~\cite{Englert,Schwinger,Scully}. This problem requires that, when the two paths of a matter-wave interferometer recombine, not only their classical positions, momenta, and all rotational degrees of freedom of the rigid body (including the three Euler angles associated with the NV-spin) coincide, but also that their quantum wave packets overlap substantially. To observe significant spin contrast in a matter-wave interferometer, we need gyroscopic stability provided by the nanodiamond's external rotation along the NV axis, as seen in \cite{Zhou:2024pdl,Rizaldy2025_RotationalStability}. These papers address both the Humpty Dumpty and the Einstein–de Haas problems by introducing an initial rotation in an NV nanodiamond.

{We extend the SGI framework by including more realistic physical effects and relaxing several idealized assumptions from previous studies, thereby better connecting simplified theoretical models with experimentally relevant conditions. The main contributions are:
\begin{itemize}
    \item We generalize the configuration of Ref.~\cite{Zhou:2024voj}, which restricts the motion to the $x$-direction, by including full two-dimensional dynamics in the $x$--$y$ plane. This enables us to capture the coupled translational and rotational motion arising from realistic effects such as magnetic-field fluctuations and IHP instabilities, thereby realizing a two-dimensional matter-wave interferometer. To stabilize oscillations along the y-axis, we apply the diamagnetic trap chip design proposed by \cite{Elahi:2024dbb}.
    \item We incorporate a lower external bias magnetic field ($B_{0} = \{0.001,0.1\} \, \text{T}$) respectively for HP and IHP stages, into the SGI scheme for nanodiamonds with triplet spin states~\cite{Doherty_2013}, explicitly accounting for its role in spin quantization, suppression of Larmor precession~\cite{paraniak2021quantumdynamicalsimulationtransversal}, Zeeman splitting~\cite{wang2025simultaneousdeterminationlocalmagnetic}, and enhanced quantum sensing sensitivity~\cite{Homrighausen}. In contrast to the idealized high-field values used in~\cite{Zhou:2024voj} ($B_{0} = \{0, 10\} \, \text{T}$), we adopt experimentally feasible conditions following~\cite{Moorthy:2025bpz}. However, the presence of $B_0$ in stage~1 introduces an asymmetry in the trajectory, causing both SGI arms to deviate significantly from their initial position and eventually exit the inverted-harmonic potential range upon entering stage~2. To address this, we modify the nonlinear magnetic field by shifting its midpoint by a distance of $2B_0/\eta$, ensuring that the SGI paths remain close to their initial position.
    \item We extend the analysis to the case of a rapidly rotating nanodiamond. Building on earlier works~\cite{Zhou:2024pdl,Rizaldy2025_RotationalStability}, we investigated the resulting spin contrast under combined translational and rotational dynamics, providing a more complete and experimentally relevant description of the HP--IHP scheme. Here we will also examine how the application of the trap will affect the rotational dynamics of the nanodiamond.
\end{itemize}
}

The structure of this paper is organized as follows. Section \ref{sec:spatial_hamiltonian} presents the Hamiltonian of the spatial motion of the NV nanodiamond in two dimensions (along the $x$- and $y$-axes) using both harmonic and inverted harmonic oscillators. 
Section \ref{sec:magnetic} discusses the magnetic field profile employed in this study, together with its specifications, which generate either a harmonic potential (linear magnetic field) or an inverted harmonic potential (non-linear/quadratic magnetic field). 
Furthermore, the transitions of the magnetic field are divided into five stages, and we assume that the transitions are smooth, which is justified by the rapid, smooth switching of the superconducting current-carrying wires we envisage for the experiment, see~\cite{Elahi:2024dbb,Xiang:2026kwd}.

In Section \ref{sec:EoM}, we formulate the two-dimensional equations of motion based on the magnetic field applied at each stage. Section \ref{sec:rotation} incorporates the Euler angle dynamics into the NV nanodiamond Hamiltonian and derives the equations of motion for libration, precession, and rotation. Section \ref{sec:wave_packet} constructs the wave packet that represents spatial and rotational motion. Finally, the last section provides the conclusions of the study.
\begin{figure*}
    \centering
    \includegraphics[width=1\linewidth]{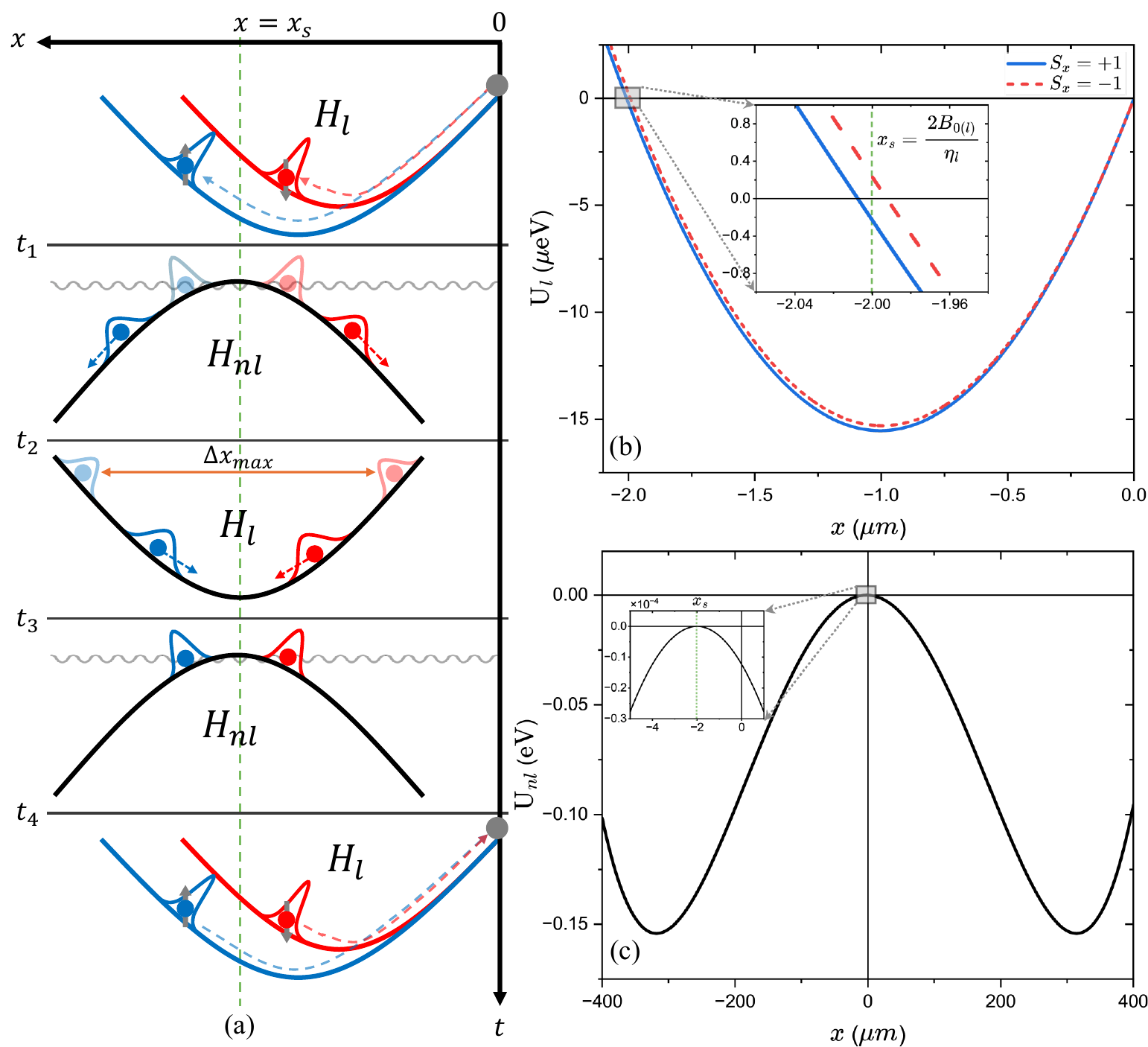}
    \caption{{(a) Illustration of the SGI protocol scheme (inspired by \cite{Zhou:2024voj,Braccini2024ExponentialCats}) in five stages (not in the real scale) corresponding to the time transitions $t_1 - t_5$ of two paths: $\ket{-1}_{(1)}\rightarrow\ket{0}_{(2,3,4)}  \rightarrow\ket{-1}_{(5)}$ and $\ket{+1}_{(1)}\rightarrow\ket{0}_{(2,3,4)}  \rightarrow\ket{+1}_{(5)}$. Each stage represents two types of Hamiltonians. The first is the Harmonic Oscillator ($H_l$), generated from a linear magnetic field profile, which governs the separation (stage 1), return (stage 3), where maximum superposition is achieved, and recombination (stage 5) of spin-up (blue line) and spin-down (red line) states. The second is the Inverted Harmonic Oscillator ($H_{nl}$), from a non-linear magnetic field profile, which functions as the enhancement and deceleration mode (stage 4). It should be noted that in this case a magnetic field bias is applied across stages, resulting in asymmetric trajectories of the two SGI arms, unlike the symmetric trajectories reported in Ref.~\cite{Zhou:2024voj}. However, we can maintain symmetrical trajectories in stages 2–4 by shifting the equilibrium position of the magnetic‑field profile to ($x_s=-2B_{0(l)}/\eta_l$), this also prevents the large displacement of both SGI arms, keeping their motion confined within the inverted‑harmonic region. (b) and (c) show the potential energy of the Harmonic Oscillator and the Inverted Harmonic Oscillator, respectively. The simulation uses NV-diamond mass $m=10^{-15}\,\text{kg}$, with a linear magnetic field bias $B_{0(l)} = 0.001\,\text{T}$ and a non-linear magnetic field bias $B_{0(nl)} = 0.1\,\text{T}$, under gradient field specifications as listed in Table~\ref{tab:magnetic_field_parameters}.}}
    \label{fig:0}
\end{figure*}

\section{Hamiltonian}
\label{sec:spatial_hamiltonian}
Before delving into the application of the Harmonic Potential (HP) and Inverted Harmonic Potential (IHP) within the Stern-Gerlach Interferometer (SGI) superposition scheme~\cite{Amit2019,Folman2013,Folman2018,doi:10.1126/sciadv.abg2879}, we briefly summarise the new results of this paper.
\begin{enumerate}
	\item {The spatial trajectory is constructed using the formalism~\cite{Zhou:2024voj}, employing linear and quadratic magnetic fields to derive the equations of motion for the harmonic oscillator (HO) and the inverted harmonic oscillator (IHO), respectively. Here, we try to give a realistic value of the bias magnetic field in all stages, as suggested by Ref \cite{Moorthy:2025bpz}. In the previous paper~\cite{Zhou:2024voj}, the bias magnetic field was too large to accommodate within a chip version of levitating the nanodiamond by Ref ~\cite{Elahi:2024dbb}. If we insist to make superposition more larger, the presence of a magnetic field bias will cause the trajectory to deviate further from its initial point, potentially moving it outside the IHP curve range. To address this issue, we modify the non-linear magnetic field equation in stages 2 and 4 by shifting its initial position by a distance of $2B_0/\eta$ from origin position.}
    \item The NV-spin transitions at each stage are as follows: in stage 1, the left and right arms of the SGI are separated into the spin states $\ket{+1}$ and $\ket{-1}$. In stages 2 to 4, the spin states $\ket{\pm 1}$ are transitioned to $\ket{0}$ using pulsed microwave. Finally, during the recombination stage, the spin states transition from $\ket{0}$ to $\ket{\pm 1}$ in both arms of the SGI. These steps are similar to those of \cite{Zhou:2024voj,Moorthy:2025bpz}.
	\item To accommodate Maxwell’s equations, we include the two dimensions, say  $x,~y$, in the SGI trajectory and introduce a trap frequency along the $y$-direction, while keeping the $x$-direction relatively flat-potential, following~\cite{Elahi:2024dbb}. Thereby, we will be able to analyse the trajectory in both directions.

	\item Rotational dynamics are also incorporated into this scheme as our main purposes in this research, adapting the idea from~\cite{Zhou:2024pdl,japha2022role,Rizaldy2025_RotationalStability}, by applying an initial angular rotation along the NV axis, $(\Omega_0)$, as suggested in \cite{Zhou:2024pdl,Rizaldy2025_RotationalStability} to stabilise the libration angle, ($\beta$), precession angle, ($\alpha$), and rotation angle, ($\gamma$). 
\end{enumerate}

The magnetic part of the Hamiltonian for the NV nanodiamond is as follows~\cite{Marshman2022, Pedernales:2020nmf}:
\begin{align}
    \hat{\textbf{H}} = \frac{\hat{\textbf{P}}^2}{2m} - \frac{\chi_\rho m}{2\mu_0}\hat{\textbf{B}}^2 +\mu(\hat{\textbf{S}}\cdot\textbf{B})+\hbar D \hat{S}^2_z + H_{trap},
    \label{eq:H_general}
\end{align}
The expression ${\hat{\mathbf{P}}^2}/{2m}$ represents the kinetic energy due to translational motion, where $\hat{\mathbf{P}}$ is the operator of translational momentum. The following term, $- {\chi_\rho m}/{(2\mu_0)}\mathbf{{B}}^2$, refers to the magnetic energy of a diamagnetic material, in this case a diamond, which has a mass susceptibility of $\chi_\rho =-6.2\times10^{-9} \text{m}^3/\text{kg}$. The third term, $\mu(\hat{\mathbf{S}}\cdot\mathbf{B})$, describes the interaction between the magnetic moment of the spin of the NV electron $\hat{\mathbf{S}}$ and the external magnetic field $\mathbf{B}$, known as the Zeeman effect. In this case, $\mu = h\times 2.8\times 10^{10} \ \text{Hz/T}$ is the magnetic moment of the NV spin. Next, $D\hat{S}_z^2$ signifies the zero-field split of the NV spin, with $D = 2\pi\times 2.8 \ \text{GHz}$ in this case. 

Lastly, we have the Hamiltonian of the harmonic trap in the $x$ and $y$ directions, or we can formulate it as $m\omega_q^2/2$, with $q=\{x,~y\}$. Note that in this scheme, we exclude the potential energy from gravity (which would affect the z-axis). This is because a strong diamagnetic field can trap and counteract the motion caused by Earth's gravity along the $z$-axis, as suggested in ref. \cite{Elahi:2024dbb}.

\section{Magnetic Field Profiles}
\label{sec:magnetic}
The magnetic fields used in this scheme are of two types: linear magnetic fields for stages 1, 3, and 5, and non-linear magnetic fields for stages 2 and 4, see the illustrative figures of these stages in \ref{fig:0}. For illustration purposes, we suppress the $y$-direction in these plots. We will discuss the motion along the $y$-direction later. The primary aspect of spatial superposition is that it being created along the $x$-direction. However, we will introduce an additional dimension, $y$, for more realistic modeling. For the linear part, we used the magnetic field~\cite{Marshman2022}:
\begin{align}
    \textbf{B}_l(x,y) = (B_{0(l)}+\eta_l x )\hat{e}_x - \eta_l y \hat{e}_y,
    \label{eq:linear_magnetic_field}
\end{align}
where $B_{0(l)}$ represents the bias magnetic field along the x-direction, and $\eta_l$ represents the gradient magnetic field, with units of  T and T/m, respectively. We wish to align the NV spin along the $x$-direction to maximize the force along this direction.

For non-linear magnetic fields \cite{Zhou:2024voj}:
\begin{align}
    \textbf{B}_{nl} (x,y) = [B_{0(nl)}-\eta_{nl} (x^2 - y^2)] \hat{e}_x + 2\eta_{nl} x y \ \hat{e}_y,
    \label{eq:non-linear_magnetic_field}
\end{align}
where $\eta_{nl}$ represents the gradient of the non-linear magnetic field with the unit $T/m^2$. The mechanism for switching on and off the two magnetic fields at each stage can be written as follows:
{
\begin{align}
    \textbf{B}(x,y) =
    \begin{cases}
        \textbf{B}_l (x,y) & \text{for } \ t \leq t_1 \\
        \textbf{B}_{nl}(x-x_s,y) & \text{for } \ t_1 \leq t \leq t_2 \\
        \textbf{B}_l(x-x_s,y) & \text{for } \ t_2 \leq t \leq t_3 \\
        \textbf{B}_{nl}(x-x_s,y) & \text{for } \ t_3 \leq t \leq t_4 \\
        \textbf{B}_l(x,y) & \text{for } \ t \geq t_5 \\
    \end{cases}
\end{align}
}
where $x_s=-2B_{0(l)}/\eta_l$ and $t_1 - t_5$ represents the time evolution at each stage, where the magnetic field in the Hamiltonian switches from linear to non-linear and vise versa. Next, we will determine the equations of motion for the x and y directions and their solutions, which will be approached analytically. 
The switch function used in this scheme for the parameters $B_0$, $\eta_l$, $\eta_{nl}$ and $t_1 - t_5$ used in this scheme can be found in Table \ref{tab:magnetic_field_parameters}~\footnote{In principle, we can provide a switching function on-and-off by refined modeling, see~\cite{Marshman2022}. First, it does not affect the trajectories or any aspects of the conclusion, and with superconducting wires, the switching can be done swiftly; therefore, for this paper we assume that the switching can be assumed as a step function. }

Fig.~\ref{fig:0} gives the illustration of all five stages, stages 1, 3, 5 are governed by a harmonic oscillator (these stages arise from the linear magnetic field profile), and stages 2 and 4 are inverted harmonic potentials (these stages arise from non-linear magnetic field profile). We show the potentials. In all these examples, we take the mass of the spherical nanodiamond to be $m\sim 10^{-15}$~kg. We show the left and right trajectories by red and blue wavepackets. We begin with the Gaussian wavepackets and the expansion of the individual wavepackets has been discussed in section \ref{sec:wave_packet}. The red and blue arrows depict their movement, and in stage-5 the two trajectories meet. We will show the numerical simulations of the trajectories in detail in section \ref{numerics}.

\begin{table}
\caption{{Here, we set the parameters used for the magnetic field profile for each stages, including the magnetic bias ($B_0$), linear magnetic gradient ($\eta_l$), non-linear magnetic gradient ($\eta_{nl}$), as well as the transition times at each stage, where $t_1, t_2, t_3, t_4$ denote stages -1, 2, 3, 4, (see Appendix \ref{appendix:analitical_solution} for further explanation about the time transition value). the levitation and superposition are created roughly at a distance $d\sim {\cal O}(10) {\rm \mu m}$ away from the chip, see~\cite{Elahi:2024dbb}.} }
\label{tab:magnetic_field_parameters}
\begin{tabular}{ccccc}
\hline
\multicolumn{1}{l}{\multirow{2}{*}{\textbf{Stages}}} & \multicolumn{4}{c}{\textbf{Parameters}}                            \\ \cline{2-5} 
\multicolumn{1}{l}{}                                 & $B_0$ (T) & $\eta_l$(T/m)          & $\eta_{nl}$($\text{T/m}^2$) & $t_{stage}$(s)            \\ \hline
1                                                    & 0.001         & 1000            & -                  & $\pi/\omega_l$         \\
2                                                    & 0.1        & -               & $10^6$             & $t_1 + \tau_2$    \\
3                                                    & 0         & $1000$ & -                  & $t_2 + \tau_3$ \\
4                                                    & 0.1        & - & $10^6$             & $t_3  + \tau_4$   \\
5                                                    & 0.001         & 1000         & -                  & $t_4+ \pi/\omega_l$  \\ \hline
\end{tabular}
\end{table}

\begin{figure*}
    \centering
    \includegraphics[width=0.85\linewidth]{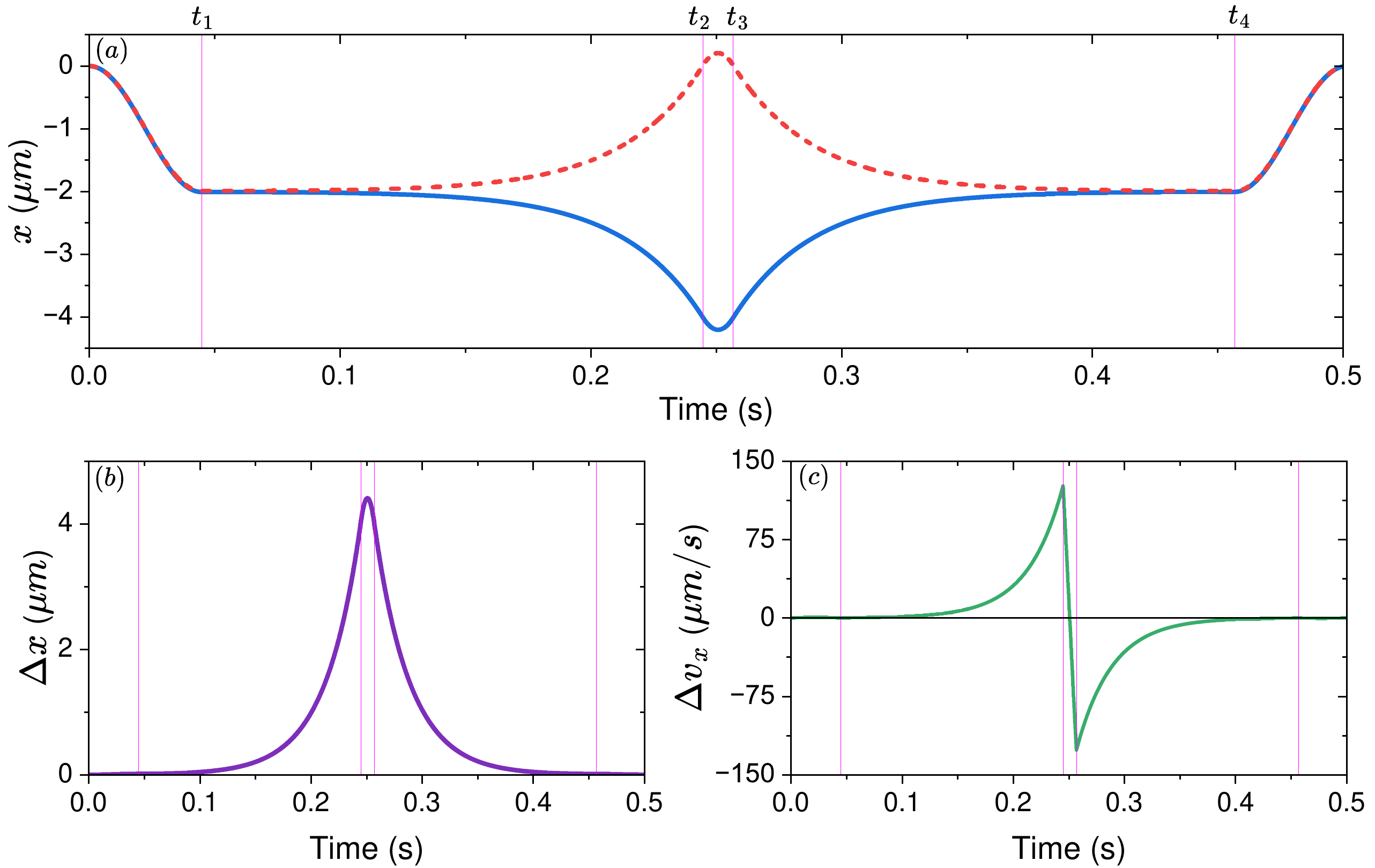}
    \caption{{The numerical results from the five-stage SGI on the x-axis, where the NV nanodiamond used has a mass of $m=10^{-15}$ kg, using magnetic field parameters according to Table 1, the purple vertical lines represent the time transition of two paths: $\ket{-1}_{(1)}\rightarrow\ket{0}_{(2,3,4)}  \rightarrow\ket{-1}_{(5)}$ and $\ket{+1}_{(1)}\rightarrow\ket{0}_{(2,3,4)}  \rightarrow\ket{+1}_{(5)}$ with time durations: 0.0447259, 0.244726, 0.2567, 0.4567 and 0.501426 s, respectively in every stage. (a) The blue and red lines represent the Left and Right SGI arms, respectively. As we can see, the presence of a magnetic field bias in the harmonic oscillator stage ($B_{0(l)} = 0.001\,\text{T}$) results in asymmetric trajectories of the two arms (In contrast, if $B_{0(l)} = 0$, the trajectories of both arms become symmetric, as shown in Ref.~\cite{Zhou:2024voj}. This asymmetric trajectories similar with Ref. \cite{Marshman2022}). Nevertheless, the presence of $B_{0(l)}$ does not affect the size of the superposition. (b) represents the superposition size (purple line) of the two SGI arms with a maximum separation value of $\Delta x_{max} \approx 4.4 \ \mu m$. (c) The green line represents the velocity difference between the two trajectories. The results have decreased significantly compared to Ref. \cite{Zhou:2024voj}, which has a superposition size of around $50 \, \mu m$. This is because, in this study, the bias value of the non-linear magnetic field (in stages 3 and 4) was reduced to a more realistic value, $B_{0(nl)} = 0.1$ T, which results in the SGI time until both arms close being twice as long as in Ref. \cite{Zhou:2024voj}, approximately $0.5$ s.}}
    \label{fig:2}
\end{figure*}

\begin{figure*}
    \centering
    \includegraphics[width=1\linewidth]{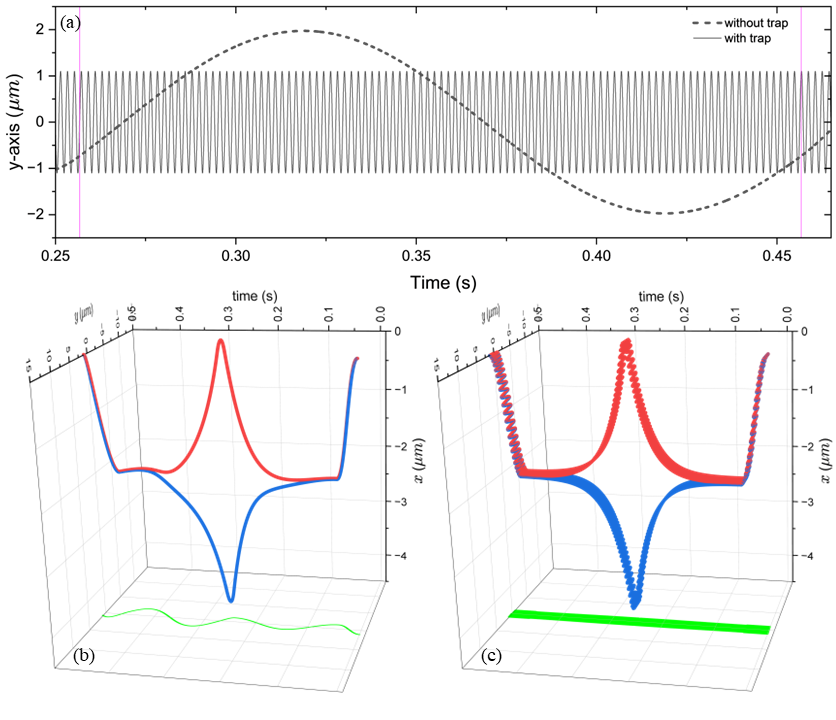}
    \caption{{(a) The numerical results of the five-stage SGI in the y-direction with $y_0 = 1.1 \ \mu m$, where the NV nanodiamond used has a mass of $m=10^{-15}$ kg, the magnetic profile parameters according to Table 1. The gray vertical lines represent the time transition of two paths: $\ket{-1}_{(1)}\rightarrow\ket{0}_{(2,3,4)}  \rightarrow\ket{-1}_{(5)}$ and $\ket{+1}_{(1)}\rightarrow\ket{0}_{(2,3,4)}  \rightarrow\ket{+1}_{(5)}$ with time durations: 0.0447259, 0.244726, 0.2567, 0.4567 and 0.501426 s, respectively in every stage. the black dashed and solid line represent without and with the given trap frequency ($\omega_y = 2 \pi \times 521$ Hz) from Ref. \cite{Elahi:2024dbb}. 
    In (b) and (c), the spatial two-dimensional motion diagrams of the nanodiamond are displayed, representing the conditions before and after the application of the trapping frequency along the y‑axis, respectively.}
    }
    \label{Fig:3}
\end{figure*}


\section{Spatial Equations of Motion}
\label{sec:EoM}
Before we formulate the equation of motion, we  first derive the Hamiltonian operator equation for positions $x$ and $y$ by substituting the linear magnetic field \eqref{eq:linear_magnetic_field}  into Eq. \eqref{eq:H_general}, so that:
\begin{align}
    \hat{{H}}_{(l)} = & \ \frac{\hat{P}_x^2}{2m} + \frac{\hat{P}_y^2}{2m} + \frac{m}{2} (\omega_l^2 + \omega_x^2)\hat{x}^2 + \frac{m}{2} (\omega_l^2 + \omega_y^2)\hat{y}^2 \nonumber \\ 
    & +\left( \mu \eta_l \hat{S}_x  - \frac{\chi_\rho m}{\mu_0} B_{0(l)} \eta_l \right) \hat{x} - (\mu \eta_l \hat{S}_y)\hat{y} \nonumber \\ 
    & - \frac{\chi_\rho m}{\mu_0} B_{0(l)}^2 + \hat{S}_x \mu B_{0(l)} + \hbar D \hat{S}_z^2
    \label{eq:Hamiltonian_linear}
\end{align}
where $\omega_l = \sqrt{-\chi_\rho\eta_l^2/\mu_0}$ represents the frequency of the harmonic potential in the case of a linear magnetic field and $\omega_\xi$ with $\xi = \{x,y\}$ is the trap frequency of the trapping potential in both $x$ and $y$ directions, with $\omega_x \ll \omega_y$, since we wish to primarily create the superposition along the $x$ direction \cite{marshman2024entanglement,DUrso16_GM,Elahi:2024dbb}.

Next, for the non-linear magnetic field Eq. \eqref{eq:non-linear_magnetic_field}, we can construct this by performing a mapping of the electronic spin to the nearest nuclear spin from the NV center, see~\cite{Abobeih2018}. This operation enhances the spin coherence and also allows us to create a simple non-linear magnetic field profile for us. Hence, we can switch off the $\hat{\textbf{S}} \cdot \textbf{B}$ term in stages involving a non-linear magnetic field. As we shall see, the switching on of the non-linear magnetic field is intended to provide enhancement of spatial superposition in stage-3 and deceleration in stage-4. Furthermore, we obtain a simplified Hamiltonian equation for a non-linear magnetic field as follows (see Appendix \ref{appendix:non-linear_Hamiltonian}):
{
\begin{align}
    \hat{H}_{(nl)} = & \ \frac{\hat{{P}}^2_x}{2m} + \frac{\hat{{P}}^2_y}{2m} - \frac{m}{2} (\omega_{nl}^2 - \omega_{x}^2)(\hat{x}-x_s)^2 \nonumber \\
    &+ \frac{m}{2} (\omega_{nl}^2 + \omega_{y}^2)\hat{y}^2   
      - \frac{\chi_\rho m B_{0(nl)}^2}{2\mu_0} + \hbar D \hat{S}^2_z,
    \label{eq:Hamiltonian_non-linear}
\end{align}
}
\begin{figure*}
    \centering
    \includegraphics[width=1\linewidth]{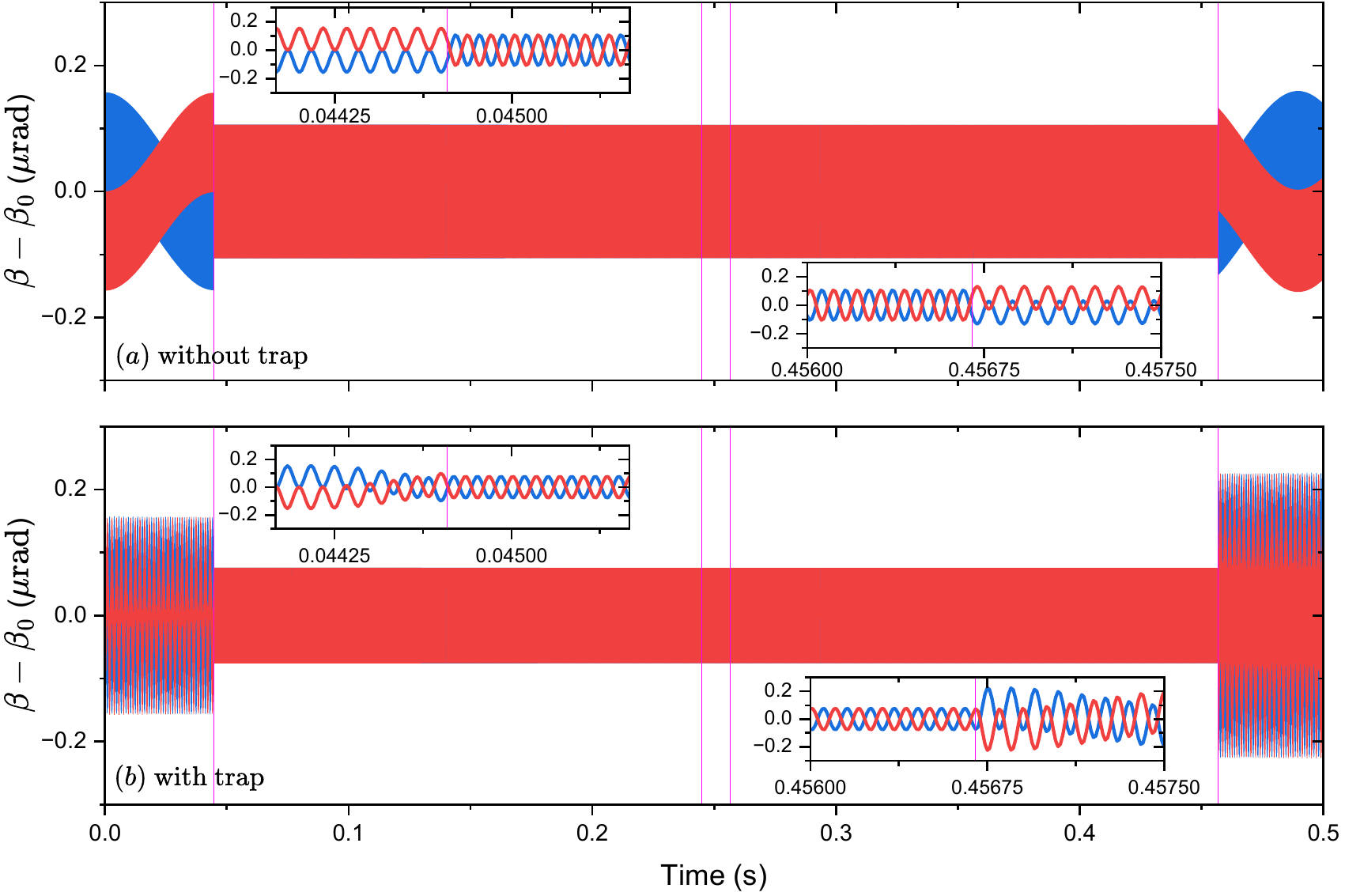}
    \caption{{The numerical solution of libration mode, Eq. \eqref{eq:beta}, uses the initial angle $\beta_0 = 0.01 \ \text{rad}$ and initial angular velocity $\Omega_0 = 2\pi \times 10^4 \ \text{Hz}$. The shape of the nanodiamond is assumed to be a perfect solid sphere with mass $m=10^{-15}$ kg and radius $R \approx 0.408 \ \mu\text{m}$. The position of the NV-spin is off-center by $d=10$ nm with a fixed angle $\alpha'=\pi/6$ from the center of mass of the nanodiamond. The magnetic field profile parameters can be found in Table 1. the purple vertical lines represent the time transition of two paths: $\ket{-1}_{(1)}\rightarrow\ket{0}_{(2,3,4)}  \rightarrow\ket{-1}_{(5)}$ and $\ket{+1}_{(1)}\rightarrow\ket{0}_{(2,3,4)}  \rightarrow\ket{+1}_{(5)}$ with time durations: 0.0447259, 0.244726, 0.2567, 0.4567 and 0.501426, respectively in every stage. The libration mode of the nanodiamond with the solid red and blue lines representing the Left and Right arms of the SGI, respectively. (a) represent without frequency trap at y-direction and (b) with frequency trap using specification of $\omega_y = 2 \pi \times 521$ Hz (Ref. \cite{Elahi:2024dbb}). One can see that the libration angle is quite stable due to the relatively high mass of the nanodiamond, and both SGI arms become more stable if the trap frequency is included.}}
    \label{Fig:4}
\end{figure*}
where $\omega_{nl} = \sqrt{-2\chi_\rho B_{0(nl)}\eta_{nl}/\mu_0}$. 

Now, we can calculate the expectation values using the Heisenberg picture formalism~\footnote{The expectation value of an operator \( \hat{A} \) satisfies the equation of motion as follows ${d\langle \hat{A} \rangle}/{dt} = {i} \langle{[\hat{H},\hat{A}]}\rangle/{\hbar}$ in the Heisenberg picture, and in this case $\hat{A}$ refers to the position operators $\hat{x}$ and $\hat{y}$.}. To simplify the calculation of the classical trajectory, we assume that the spin direction occurs only along the $x$-axis. This is admittedly an idealized approach because the Larmor precession can cause the spin to tilt towards the other axis. To avoid that, we align the bias magnetic field $B_0$ along the $x$-direction. {Moreover, approximately in this scheme, $\omega_l \sim \omega_{nl} \approx 10 \, \text{Hz}$ compared to the trap frequency, see ref. \cite{Elahi:2024dbb}, $\omega_x/2\pi \approx 10^{-2} \, \text{Hz}$, so that $ \omega_{nl} \sim \omega_l \gg \omega_x$. However, along the $y$ direction, $\omega_y >\omega_{nl}\sim \omega_y/2\pi \approx 10^2 \, \text{Hz}$}
\footnote{{If we refer to the trap frequencies generated in the reference \cite{Elahi:2024dbb} from the current carrying chip design, after coordinate adjustment in this case, where $\omega_x = 2 \pi \times 0.01$ Hz and $\omega_y = 2 \pi \times 521$ Hz, compared to $\omega_{nl} = 31.412$ Hz and $\omega_l = 70.241$ Hz, so it will always be $\omega_y > \omega_{nl} > \omega_l$.}}. 

We can now obtain the acceleration equations for the linear part of the magnetic field, i.e., stages 1, 3, 5, as follows :
\begin{align}
    \frac{d^2 \langle x \rangle_l}{dt^2} & = -\omega_l^2 \langle x \rangle_l - \frac{s\mu \eta_l}{m} +\frac{\chi_\rho}{\mu_0} B_{0(l)} \eta_l, \label{eq:x_linear} \\
    \frac{d^2 \langle y \rangle_l}{dt^2} & = -(\omega_l^2+\omega_y^2) \langle y \rangle_l \label{eq:y_linear} ,
\end{align}
where $s=\{-1, +1\}$. 

For the non-linear part of the magnetic field, i.e., stages 2, 4, we have equations of motion:
{
\begin{align}
    \frac{d^2 \langle x \rangle_{nl}}{dt^2} & = \omega_{nl}^2 (\langle x \rangle_{nl} -x_s), \label{eq:x_non-linear} \\
    \frac{d^2 \langle y \rangle_{nl}}{dt^2} & = -(\omega_{nl}^2+\omega_y^2) \langle y \rangle_{nl} \label{eq:y_non-linear},
\end{align}
}
\begin{figure*}
    \centering
    \includegraphics[width=1\linewidth]{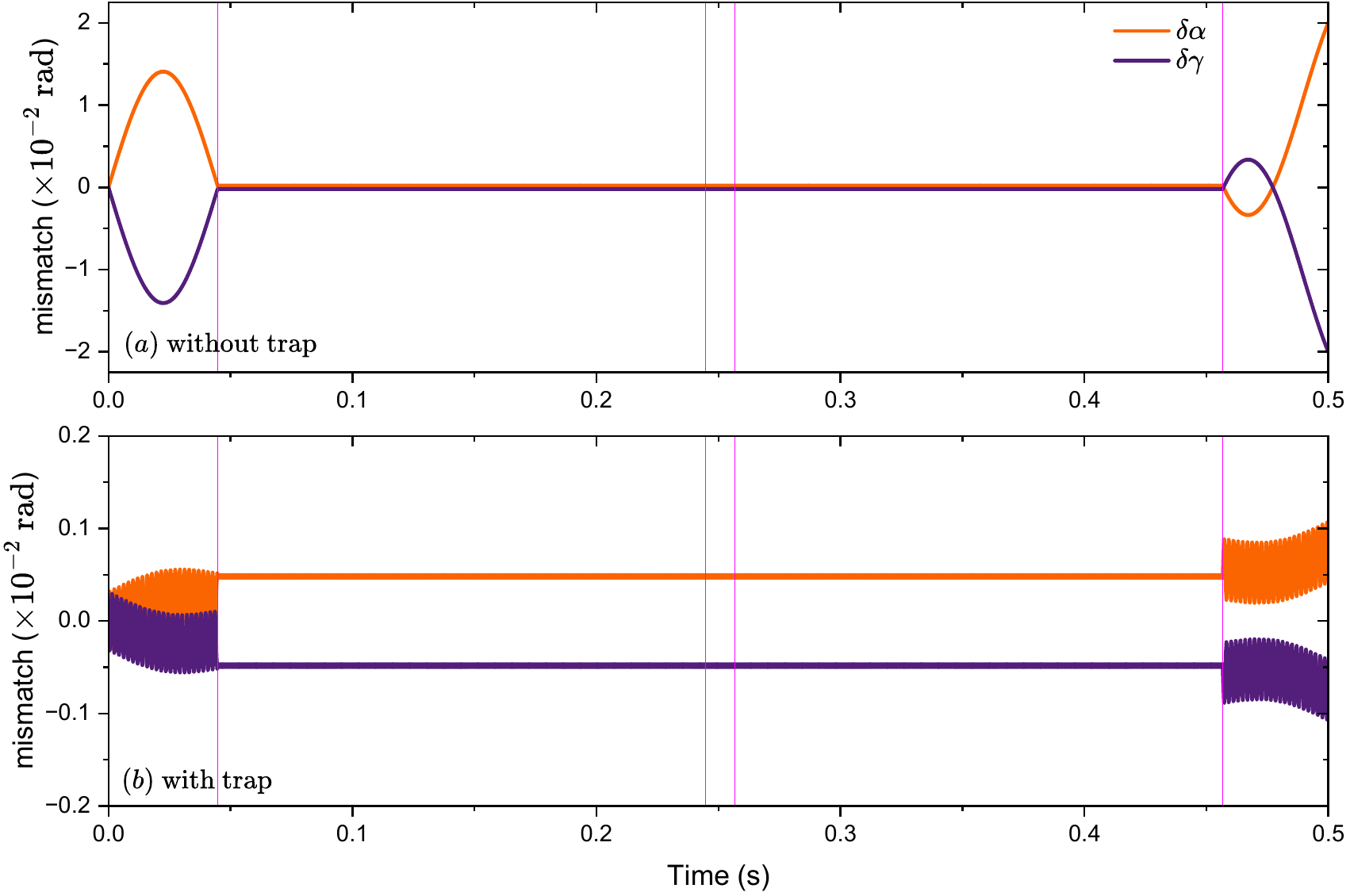}
    \caption{{The numerical solution of Eq. \eqref{eq:alpha} and \eqref{eq:gamma} use the initial angle $\beta_0 = 0.01^\text{o}$ and initial angular velocity $\Omega_0 = 2\pi \times 10^4 \ \text{Hz}$. The shape of the nanodiamond is assumed to be a perfect solid sphere with mass $m=10^{-15}$ kg and radius $R \approx 0.408 \ \mu\text{m}$. The position of the NV-spin is off-center by $d=10$ nm with fix angle $\alpha'=\pi/6$ from the center of mass of the nanodiamond. The magnetic field profile parameters can be found in Table 1. The purple vertical lines represent the time transition of two paths: $\ket{-1}_{(1)}\rightarrow\ket{0}_{(2,3,4)}  \rightarrow\ket{-1}_{(5)}$ and $\ket{+1}_{(1)}\rightarrow\ket{0}_{(2,3,4)}  \rightarrow\ket{+1}_{(5)}$ with time durations: 0.0447259, 0.244726, 0.2567, 0.4567 and 0.501426 s, respectively in every stage. The mismatch (between left and right SGI arms, $\delta q = q_L - q_R, \ q=\{\alpha,\gamma\}$) of the precession angle ($\alpha$) in brown solid line and the rotation angle ($\gamma$) in purple solid line, where both have a symmetrical all over the stages ($\delta\alpha \approx -\delta\gamma$) for both cases, (a) represent without frequency trap at y-direction and (b) with frequency trap using spesification of $\omega_y = 2 \pi \times 521$ Hz (Ref. \cite{Elahi:2024dbb}).}}
    \label{Fig:5}
\end{figure*}

The derivations of the solutions for these equations of motion, from a linear and non-linear magnetic field, can be found in the appendix \ref{appendix:analitical_solution}. We will follow \cite{Moorthy:2025bpz}, which includes more realistic parameters for the magnetic field at each stage, although the trade-off is a significantly smaller superposition size ($\sim 1 \mu m$) over a complete loop time of $t=0.3$ s, three times longer than the scheme of Ref.~\cite{Zhou:2024voj}, which has a superposition size of $\sim 50 \ \mu m$). 
{The presence of magnetic-field bias causes the two trajectories to lose their symmetry and pushes them far beyond the inverted-potential boundary into the quartic region (see Fig.~\ref{fig:0}(c)). To address this issue, in stage~1 both trajectories are released until time $t = \pi/\omega_l$, when they reach their maximum separation and their instantaneous velocities become zero. At that moment, the positions of the two trajectories are located around the equilibrium point $x_s = -2B_{0(l)}/\omega_l$. This equilibrium point is then used to shift the magnetic-field profile in stages~2–4 so that the two arms of the SGI become symmetric.
This improvement become massive difference to Ref. \cite{Moorthy:2025bpz}, instead of stopping stage 1, the transition occurs at the maximum speed in the end of stage 1, around $t_1 = \pi/(2\omega_l)$ to improve the superposition size in stage 3. Also, we set the magnetic gradient value in the non-linear stage (stages 2 and 4) to $\eta_{nl} = 10^6 \ \text{T/m}^2$ to increase the superposition size (such a value of the magnetic field gradient is feasible, see~\cite{Elahi:2024dbb}).}

\section{Numerical Results for spatial superposition}
\label{numerics}

Fig.\ref{fig:2} illustrates the spatial evolution of the five stages of the SGI based on the potentials shown in Fig.\ref{fig:0} (a). The trajectories of both left and right arms of the interferometer are depicted in blue and red dashed curves. {We can see that the trajectories are asymmetric on stage 1 and 5 and maintai its symmetrical path at stage 2-4 around $x_s$}. The stage-1 separates the left and right trajectories due to the $\propto \pm S\cdot \eta_l $ contribution, we can see this already in Fig.\ref{fig:0} (a) during stage $t_1$. During stage-2, the superposition size is enhanced. {We switch to stage-2 when the left and right trajectores became 0 m/s and activated non-linear magnetic field with shifted position to $x_s$}. Figs.~\ref{fig:2}(b) and (c), show the superposition size, which in our case is $\Delta x\sim 4.4 {\rm \mu m}$ for mass $m\sim 10^{-15}$kg, and the difference in velocity. Note that we are able to close the trajectories in $t=0.5$s, which means that $\Delta x=0$, and $|\Delta v|=0$ at $t=0.5$ s. These figures have been shown for the modest bias magnetic field $B_0\sim 10^{-3}$ T, which is compatible with our superconducting current-carrying chip design to levitate the nanodiamond, see~\cite{Elahi:2024dbb}. If we levitate the nanodiamond from the chip at a distance of $d\sim {\cal O}(10) {\rm \mu m}$, we are also able to generate $\eta_l$ and $\eta_{nl}$ similar to the figures shown in Table.~\ref{tab:magnetic_field_parameters}.
The vertical lines in Figs. \ref{fig:2} (a, b, c) show the transitions between different stages, i.e., stage -1, 2, 3, 4, 5 and also the spin transitions of the NV, mapping the NV spin to the nuclear spin and vise-versa. This is a well-known procedure to enhance the coherence of the electronic spin \cite{Abobeih2018}, and to enhance the superposition size, as shown first in~\cite{Zhou:2022frl,Zhou:2022jug}.

Figs. \ref{Fig:3} (a,b,c) show the motion of the nanodiamond along the $y$-direction, with ($\omega_y= 2\pi\times 521$ Hz, which is compatible with the levitating chip design of \cite{Elahi:2024dbb}, and without the trap $\omega_y=0$ Hz. The vertical lines in \ref{Fig:3} (a) show the transition between different stages and also the mapping of electronic spin to nuclear spin and vise-versa, as also shown in 
Figs. \ref{fig:2} (a, b, c).
We notice that a tight trap can minimize the oscillations in the $y$-direction, and restricts the superposition size along the $y$-direction significantly; compare Figs.~\ref{Fig:3} (b,c). The blue curve and the red dashed lines show the left and right arms of the trajectories. Our current simulations are the first example of $2$-spatial dimensional SGI based on the magnetic field and currents compatible with the current-carrying chip design of ~\cite{Elahi:2024dbb}~\footnote{We also constructed a similar $2$-spatial dimensional SGI, but for a fixed magnet setup, see~\cite{marshman2024entanglement}, based on the magnetic field configuration of \cite{DUrso16_GM}. The advantage of our current setup is that we are able to create a significantly large spatial superposition in a shorter time scale, i.e. $\Delta x_{max}\sim 4.4 {\rm \mu m}$ in $t=0.5$s for $m\sim 10^{-15}$kg. This is mainly possible because we are using the trap frequencies of $\omega_x, \omega_y$ arising from the magnetic field profiles of a superconducting current-carrying chip design ~\cite{Elahi:2024dbb}. The magnetic field gradients are typically large in a chip version as compared to the fixed-magnet configurations.}.

The superposition size and the closure of the interferometric time are encouraging from the experimental perspective to test the spin entanglement witness to the quantum nature of gravity in a lab, see~\cite{Schut:2025blz}, we will still need to make sure that the decoherence rates are well under control within $10^{-2}-10^{-3}$Hz. We will revisit this discussion in our concluding section.

\section{Rotational dynamics of nanodiamond}
\label{sec:rotation}

We have shown earlier that imparting initial rotation on a nanodiamond along its NV axis can help stabilize the orientation and reduce oscillations in other directions, see~\cite{Zhou:2024pdl,Rizaldy2025_RotationalStability}. 
and help to maintain the quantum superposition more consistently.
The rotational part of the nanodiamond's Hamiltonian can be written as \cite{Zhou:2024pdl,Rizaldy2025_RotationalStability}~\footnote{See, \cite{Rizaldy2025_RotationalStability} for a complete derivation of the Hamiltonian.}:
\begin{align}
    \hat{\textbf{H}} = \frac{\hat{\textbf{P}}^2}{2m} + \frac{\hat{\textbf{L}}^2}{2I} - \frac{\chi_\rho m}{2\mu_0}\hat{\textbf{B}}^2 +\mu(\hat{\textbf{S}}\cdot\textbf{B})+\hbar D \hat{S}^2_\parallel + \hat{H}_{trap},
    \label{rotHam}
\end{align}
one can see that the additional ${\hat{\textbf{L}}^2}/{2I}$ term represents the rotation of the nanodiamond, with $\hat{\textbf{L}}$ being the angular momentum, and $I$ is the moment of inertia of the nanodiamond. There are three Euler angles involved in this scheme, namely $\{\alpha,\beta,\gamma\}$ representing the precession, libration, and rotation mode, respectively. which rotates at coordinates $\{\hat{n}_1,\hat{n}_2,\hat{n}_3\}$. For our purpose, we assume that the nanodiamond is a sphere, see appendix \ref{appendix:euler_angle}, where we have shown the Euler angles and the coordinate frames.

Here, we assume that the spin orientation of the NV nanodiamond, $\widehat{n}_s = \{\cos{\beta},\sin{\beta},0\}$, is parallel to $\hat{n}_3$ the same configuration as in \cite{Zhou:2024pdl}, $S_\parallel = \textbf{s}\cdot\hat{n}_s$, where the nanodiamond rotated about the $x$-axis with initial angle $\beta_0 \approx0.01^o$. We also assume that the location of the NV spin in the nanodianond is off center with the distance from center $\Vec{d}$, with $\alpha'$ as the fix angle between $\hat{n}_3$ and $\Vec{d}$, so $\Vec{d} = d\{\cos{(\beta+\alpha')},\sin{(\beta+\alpha')},0\}$, we set $d = 10$ nm and $\alpha'=\pi/6$ rad for a nanodiamond of radius $R\approx 0.408 {\rm \ \mu m}$. In order to make the rotation gyroscopically stable, we impart the initial rotation 
$\Omega_0 \sim {\cal O} ({\rm KHz})$ range along $\widehat{n}_s$.

In the presence of rotation, we can rewrite the equations of motion of Eq. (\ref{eq:x_linear},\ref{eq:y_linear})  (see Appendix \ref{appendix:euler_angle} for the derivations), as
\begin{align}
    \frac{d^2 \langle x \rangle_l}{dt^2} & = -\omega_l^2 \langle x \rangle_l  +\frac{\chi_\rho}{\mu_0} B_{0(l)} \eta_l - \frac{s\mu \eta_l}{m} \cos{\beta}, \label{eq:x_linear-rot} \\
    \frac{d^2 \langle y \rangle_l}{dt^2} & = -(\omega_l^2+\omega_y^2) \langle y \rangle_l + \frac{s\mu \eta_l}{m} \sin{\beta}. \label{eq:y_linear-rot}
\end{align}
Note that the equation of motion for the non-linear part is still the same as in Eq. (\ref{eq:x_non-linear}, \ref{eq:y_non-linear}), due to the spin transition from $\ket{\pm 1} \rightarrow \ket{0}$ at all stages involving the non-linear parts, i.e., stages-2 and 4.

One can see that the additional term ${\hat{\textbf{L}}^2}/{2I}$ represents the rotation of the nanodiamond, see Eq.~\eqref{rotHam}, where $\hat{\textbf{L}}$ being the angular momentum, and $I$ is the moment of inertia of the nanodiamond (for a spherical nanodiamond, we have: $I=2mR^2/5$). The formulation of Hamiltonian for rotation in momentum form can be written as (see ref. \cite{Zhou:2024pdl,japha2022role,Rizaldy2025_RotationalStability})
\begin{align}
    H_{rot} =& \ \frac{p_\beta^2}{2I} + \frac{p_\gamma^2}{2I} + \frac{(p_\alpha - p_\gamma \cos{\beta})^2}{2I\sin^2{\beta}} + \mu \hat{\mathbf{S}} \cdot \mathbf{B}(\vec{x} + \vec{d}),
    \label{spin-Ham0}
\end{align}

Using the same configuration of the two magnetic fields for both linear and non-linear cases, the classical libration mode $\beta$ for rotation is given by \cite{Zhou:2024pdl,Rizaldy2025_RotationalStability}:
\begin{align}
	\Ddot{\beta} &= \frac{(p_\alpha-p_\gamma \cos{\beta})(p_\alpha \cos{\beta} - p_\gamma)}{I^2 \sin^3{\beta}} - \frac{\mu s {B_{NV}}}{I}\nonumber \\
    &\approx -\Omega^2 (\beta - \bar{\beta}), \label{eq:beta}
\end{align}
where $p_\alpha = I\Omega_0 \cos{\beta_0}$ and $p_\gamma = I\Omega_0$. 

The magnetic field in the off-centered NV is defined as 
$$B_{NV} = -B_y\cos {\beta} -  B_x \sin{\beta}+d \eta_l \sin{(\alpha'+2\beta)}$$ and $\Omega =\sqrt{ \Omega_0^2 - {\mu s {B_x}/I}}\approx\Omega_0$ is the libration frequency, where the initial frequency is given by: $\Omega_0$, and $\bar{\beta}(t) \approx \beta_0 + \mu s B_x(t) \beta_0/(I\Omega^2_0) $ is the equilibrium position of the libration angle. 

For the precession ($\alpha(t)$) and the rotation mode ($\gamma(t)$), we have the equations of motion~\cite{Rizaldy2025_RotationalStability}:
\begin{align}
	\dot{\alpha} &= \frac{\Omega_0}{\sin^2{\beta_0}} (\cos{\beta_0} - \cos{\beta}) \approx \frac{\Omega_0}{\beta_0} (\beta-\beta_0) , \label{eq:alpha}\\
	\dot{\gamma} &= \frac{\Omega_0}{\sin^2{\beta_0}} (1- \cos{\beta_0}\cos{\beta}) \approx \Omega_0-\frac{\Omega_0}{\beta_0} (\beta-\beta_0). \label{eq:gamma}
\end{align}
Here, Eqs. \eqref{eq:alpha},~\eqref{eq:gamma} can be solved by specifying the initial conditions. In our case, we take $\beta(0) = \beta_0 = 0.01\,\text{rad}$, $\alpha(0) = \gamma(0) = 0$, and setting the initial angular velocity at $\Omega_0 = 2\pi \times 10^4$ Hz. From Eq.~\eqref{eq:beta}, we see that the libration mode always has a harmonic oscillator form, even when a quadratic (non-linear) magnetic field is used.


\section{Numerical results for the rotating nanodiamond}

Ref.~\cite{Japha:2022phw} has shown that rotation of a nanodiamond along the NV axis, known as the libration mode, in the SGI setup is inevitable due to the external torque, whether the NV center is located close to the center of mass or away from the center of mass. The libration of the nanodiamond leads to a loss in the contrast of the spin population measurements; hence, more repetition of the experiment will be required to overcome the challenges. To resolve this challenge, refs.~\cite{Zhou:2024pdl,Rizaldy2025_RotationalStability} proposed that we can impart an initial rotation $\Omega_0$ along the NV axis, which gives a gyroscopic stability, as we have argued in this paper as well. In all stages-1,2,3,4,5 gyroscopic stability helps to stabilize $\beta$. This can be seen in our numerical simulations in Fig.\ref{Fig:4}. For the purpose of illustration, we take $\beta_0=0.01$ rad, and we can see that $\beta$ evolves to become smaller
after stage-1 and remains constant in all stages, except when we map the nuclear spin to the electronic spin, when the $\beta$ evolves again. The evolution of $\beta$ is closely related to the term $S\cdot\nabla B$ in the Hamiltonian. When this term is absent, when the electronic spin is mapped to the nuclear spin, the evolution of $\beta$ freezes. At the end of the one-loop interferometer, $\beta$, for both the left and right parts of the trajectories, it tends to become even smaller than the initial value of $\beta_0$. We also note that the trapping potential in the $y$-direction has a small bearing on the evolution of $\beta$. The amplitude of the oscillations for $\beta$ is smaller compared to the case where there is no trapping in the $y$-direction. In this case, the $y$-direction has a trapped potential with a frequency $\omega_y=2\pi \times521$Hz. 

Note the following: there is an interesting phenomenon in the transition from stage 4 to stage 5 in Fig. \ref{Fig:5} where the amplitude of the libration mode becomes higher than in stage 1 (increasing by 0.08\% without the trap and 0.03\% with the trap along the y-axis). This is caused by the contribution of the initial values of the translational motion in the $x$ and $y$ directions, which differ during the transition from stage 4 to stage 5 (see Fig. \ref{fig:2}), compared to the initial conditions ($x(0)=0$ and $\dot{x}(0)=0$). The libration mode amplitude in stage 5 becomes even higher because of the coupling contribution between the libration and precession angles in the torque about the y-axis, which affects the torque on the NV nanodiamond and consequently influences the libration mode solution. However, the amplitude decreases to 0.05\% when the trap is activated. 

Similarly, Figs.\ref{Fig:5} show the evolutions of $\alpha$ and $\gamma$. In these figures, we show $\delta \alpha $ and $\delta \gamma$. From Eqs.~\eqref{eq:alpha}--\eqref{eq:gamma}, we can see that their evolution is the same, i.e. the right-hand sides remain identical. Hence, for a given $\beta_0$, the evolution of $\alpha$ and $\gamma$ follows the evolution of $\beta$. Since $\beta(t) \ll \beta_0$, changes in $\alpha$ and $\gamma$ follow a similar evolution, as can be seen from our analysis. Here, we can see that $\delta \alpha, ~\delta \gamma$ evolve during stages 1 and 5. However, $\delta \alpha,~\delta \gamma$ cease to evolve during other stages. This is because $\beta$ also does not evolve when the electronic spin is mapped to the nuclear spin. Note that for the left and right arms of the trajectories $\delta \alpha$ and $\delta \gamma$ the same remains. 


\section{Wave-Packet Evolution}
\label{sec:wave_packet}

We use a Gaussian Shaped Wave-Packet because in both harmonic and inverted harmonic potentials. The Gaussian wave retains its shape throughout the evolution of time. This allows for the derivation of analytical solutions for the width, phase, and center of the wave by simply calculating the changes in these parameters \cite{Rauh2016,Barton1986,Yuce2021,Rajeev2018,Pedrosa2015}. 
The ability to achieve minimal uncertainty states facilitates the calculation of dispersion and phase evolution for coherence and interference contrast analysis, and it simplifies the analysis of the impacts of magnetic field, magnetic field gradient fluctuations, and initial position deviation, explicitly through the Gaussian parameters. 
This approach simplifies experimental optimization and theoretical validation compared to other approaches that do not provide exact solutions in quadratic potentials. (see Ref. \cite{Zhou:2024voj} for a more comprehensive explanation and derivation). 

\begin{align}
	\psi(x,t) =& \ N(t) \exp\bigg\{-\frac{1}{4\sigma^2(t)}[x-x_c(t)]^2 \nonumber \\
    &+i\left[\frac{a(t)}{4}x^2+b(t)x+c(t)\right]\bigg\},
    \label{eq:wave-function_spatial-x}
\end{align}
where
\begin{widetext}
\begin{align}
	\sigma_x^l(t) =& \ \sigma_0\sqrt{\frac{\hbar^2}{4m^2\omega^2_l \sigma_0^2} \sin^2(\omega_l t) + \left[\frac{a_0 \hbar}{2m \omega_l} \sin(\omega_l t) +  \cos(\omega_l t)\right]^2} \label{eq:sigma_l} \\
    x_c^l(t) = & \ \frac{\hbar a_0 x_0}{2m\omega_l} \sin{(\omega_l t)}+x_0 \cos{(\omega_l t)}+\frac{\hbar b_0}{m\omega_l} \sin{(\omega_l t)} +\frac{A_0[1-\cos{(\omega_l t)}]}{m \omega_l^2} \label{eq:x_c_l} \\
    \sigma_x^{(nl)}(t) =& \ \sigma_0\sqrt{\frac{\hbar^2}{4m^2\omega^2_{nl} \sigma_0^2} \sinh^2(\omega_{nl} t) + \left[\frac{a_0 \hbar}{22m \omega_l} \sinh(\omega_{nl} t) +  \cosh(\omega_{nl} t)\right]^2} \label{eq:sigma_nl}\\
    x_c^{(nl)}(t) = & \ \frac{\hbar a_0 x_0}{2m\omega_l} \sinh{(\omega_{nl} t)}+x_0 \cosh{(\omega_{nl} t)}+\frac{\hbar b_0}{m\omega_{nl}} \sinh{(\omega_{nl} t)}. \label{eq:x_c_nl}
\end{align}
\end{widetext}
The upper subscript $l$ represents the linear magnetic field, valid during the stages where the magnetic field is linear (stages 1, 3, and 5). Conversely, in the stages where the magnetic field is non-linear (stages 2 and 4), we have $\omega_l \rightarrow \omega_{nl}$, and the trigonometric functions transform into hyperbolic functions: $\sin(\omega_l t) \rightarrow \sinh{(\omega_{nl} t)},$ and $\cos(\omega_l t) \rightarrow \cosh{(\omega_{nl} t)}.$ 

In the last term of Eq. \eqref{eq:x_c_l}, there is an additional contribution arising from the presence of a magnetic field bias in the Hamiltonian harmonic oscillator, $A_0 = S_x \mu-\chi_\rho m B_{0(l)} \eta_l/\mu_0$. However, unlike the Hamiltonian harmonic oscillator in Eq. \eqref{eq:Hamiltonian_linear}, which has two independent variable terms ($\hat{x}^2$ and $\hat{x}$), the Hamiltonian inverted harmonic oscillator in Eq. \eqref{eq:Hamiltonian_non-linear} contains only the independent variable $\hat{x}^2$. Therefore, there is no parameter as $A_0$ appears in the solution in Eq. \eqref{eq:x_c_nl}. Its complete definition can be found in Appendix~\ref{appendix:Spatial Contrast}, together with the parameters $N(t)$, $a(t)$, $b(t)$ and $c(t)$. If the initial state is given by a Gaussian wave packet profile, it must be $N_0 = [{1}/{(2\pi\sigma_0^2)}]^{1/4}, a_0 = 0, b_0 = {p_0}/{\hbar}, \ \text{and} \  c_0 = -{p_0 x_0}/{\hbar}.$

\begin{figure}
    \centering
    \includegraphics[width=1\linewidth]{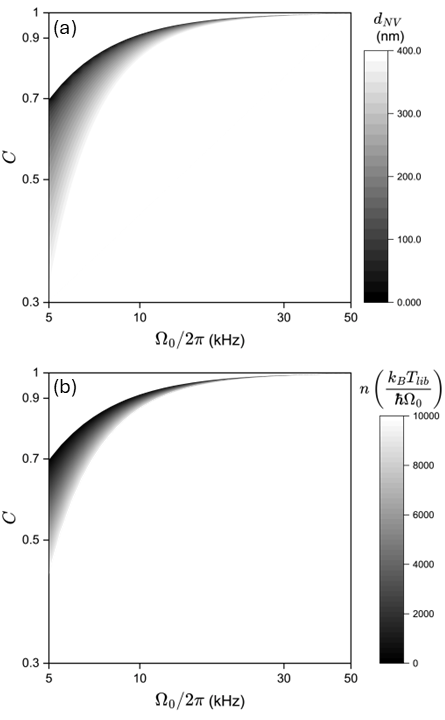}
    \caption{The contrast of the libration mode at the recombination time $(t_{5})$ for a nanodiamond of mass $10^{-15}\,\mathrm{kg}$ (see Eq. \eqref{eq:contrast_temprature}), which has a radius $R=408.58$ nm, with a magnetic bias field $B_{0}=0.001\ \mathrm{T}$, $\sigma_{p_{q}}=5\hbar$,  $\abs{\delta q} = 0.1$ rad, where $q=\{\alpha,\gamma\}$, and a magnetic field gradient $\eta_{l}=4460\ \mathrm{T}$, is shown for varying initial angular velocity. Panel (a) shows the variation of the NV off-center distance with projection angle $\alpha_{0}=\pi/6$ and occupation number $n=20$. Panel (b) shows the variation of the occupation number with the NV offset set to $d=10\ \mathrm{nm}$. One can see that the NV center's distance from the nanodiamond's center affects the contrast; this effect was neglected in Ref.~\cite{Zhou:2024pdl,Rizaldy2025_RotationalStability}, but becomes very significant here because this scheme uses a high magnetic gradient.}
    \label{fig:5}
\end{figure}

Now we can calculate the contrast of spatial motion in the x-direction, which is also the overlap of the two SGI arms $\psi_L(x,t)$ and $\psi_R(x,t)$, based on equation \eqref{eq:wave-function_spatial-x} with the definition of \cite{Schwinger} as follows:
\begin{align}
    C_x=\exp[-\frac{(x_R - x_L)^2}{8\sigma_x^2(t)}-\frac{\sigma_x^2(t)}{2}(b_R-b_L)^2]
\end{align}

Unlike the case of spatial motion, for the rotation case, the equation of motion always takes the form of a harmonic oscillator, whether the magnetic field is linear or non-linear.

Therefore, to describe the evolution of the wave function, one can use the approach employed by Ref. \cite{Zhou:2024pdl,Rizaldy2025_RotationalStability}, and we have spin contrast definition with thermal noise at recombination time ($t_5$) See Appendix \ref{appendix:humpty-dumpty}. 
\begin{align}
    C_{t h}>\exp \left[-\frac{\delta \alpha^2 \sigma_{p_\alpha}^2}{2 \hbar^2}-\frac{\delta \gamma^2 \sigma_{p_\gamma}^2}{2 \hbar^2}- \frac{16(1+2 n) \mu^2 \mathcal{B}^2}{\hbar I \Omega_0^3}\right]. \label{eq:contrast_temprature}
\end{align}
where $\mathcal{B}(t_5) = (B_{0}^{(l)}\beta_0 - y_0 \eta_l + d\eta_l \sin{\alpha_0})$ and $n = k_B T/(\hbar \omega_0)$ is defined as the occupation number, which represents thermal noise. The result of the contrast calculation can be seen in Fig. \ref{fig:5}. From Fig. \ref{fig:5}(a), it can be seen that the position of the NV-center far from the center reduces the contrast value (at recombination time $t_5$, $\abs{\delta\alpha} = \abs{\delta \gamma} \approx 0.1$ rad with trap). 
This occurs in our case because of the application of a very large magnetic-field gradient at the recombination stage ($\eta\sim5\times 10^3$ T), causing the two arms of the SGI to experience very different Zeeman torques, since their magnetic moments are oriented in opposite directions and the field fluctuates across the strong gradient. As a result, the two rotational trajectories $\beta(t)$ for the two spin states diverge significantly along the interferometer. Because the angular mismatch $\delta \beta(t_5)$ becomes much larger than the quantum coherence length, the overlap between the two coherent states decreases exponentially, so the spin-interference contrast drops to nearly zero. Fig. \ref{fig:5}(b) shows the reduction in contrast due to the presence of thermal noise. At T = 0 K (the quantum ground state), the wave packet of the libration mode is narrow and centered exactly at the equilibrium position ($\beta_0$ = 0.01 rad). Because the wave packet is narrow, when the interferometer is closed, the two wave packets still significantly overlap even though their centers are slightly shifted. For $T > 0$, thermal vibrations provide additional rotational kinetic energy to the libration mode, causing the wave packet to expand. As a result, the equilibrium position becomes smeared around its mean value. When recombination occurs, the left and right arms are more spread out and their centers are farther apart, so the overlap decreases exponentially. Nevertheless, this condition can be remedied by applying a rapid initial rotation.
\section{Conclusions}
\label{Conclusion}
Our main focus is to examine the effect of rotational and spatial degrees of freedom in the process of creating a macroscopic quantum superposition with the help of harmonic/inverted harmonic potentials created solely via the magnetic field and its gradients. We focused on the magnetic-field bias, the role of the trapping potential, and the effective potential profiles that arise at each stage of the evolution, both at the level of classical trajectories and in the description of quantum wave packets. The presence of a magnetic-field bias affects the behavior of the trajectories of the two SGI arms (similar to Ref.~\cite{Marshman2022} result), making the two trajectories asymmetric. However, the presence or absence of a magnetic-field bias does not affect the superposition size (at stage 3). Even so, the magnetic-field parameters still need to be tuned so that the two SGI arms can be closed. {For a test mass of $10^{-15}\,\mathrm{kg}$, this scheme produces a superposition size of  approximately $4.4\,\mu\mathrm{m}$. For realistic parameter choices, this represents a very significant improvement (compared to Ref.~\cite{Moorthy:2025bpz}, which used a similar scheme and obtained $\sim 1\,\mu\mathrm{m}$, and Ref.~\cite{Marshman2022}, which used a mass of $10^{-17}\,\mathrm{kg}$). 
From the QGEM perspective, based on the investigation in Ref.~\cite{Schut:2025blz}, superposition at this scale is already relevant to achieve a {negative entanglement 
witness}, especially when the minimal separation between the superposed branches can be reduced to around $35\,\mu\mathrm{m}$ and the decoherence is kept sufficiently low.}

For the rotational equations of motion, the libration equations of motion will always form a harmonic potential because the initial angular velocity is dominant, even though the system has already entered the IHP stage in its spatial motion. The gyroscopic stability of the nanoparticle is essential to mitigate and address the Humpty-Dumpty problem in the spin contrast, as shown in earlier works~\cite{Zhou:2024pdl,Rizaldy2025_RotationalStability}. Here, we reiterate the importance of imparting an initial rotation to the nanoparticle as a means of initial-state preparation.

The trapping potential in the y-direction not only stabilizes the spatial motion of the particle along $y$, but also stabilizes the libration mode when the particle rotates rapidly. This can be seen in stage 1 and 5 in Fig. \ref{Fig:4} and \ref{Fig:5}, from the comparison of the librational-mode amplitudes before and after the trap in the y-direction is applied. To formulate the wave packet, we use the same formalism as \cite{Zhou:2024voj}, namely the Gaussian-shaped wave packet equation. However, we now include the magnetic-field bias in the initial Lagrangian formulation (see Appendix \ref{appendix:WPE_general}). It can be seen that the inclusion of the magnetic-field bias does not change the evolution of the wave-packet width, but it does affect the motion of the wave-packet center. As a consequence, it takes slightly longer for the two SGI arms to completely overlap after separation, and also during recombination.

{Our analysis provides a framework that can be extended to more complex scenarios. For example, the calculations presented here provide a basis for examining various nanoparticle geometries by incorporating the spatial profile of the moment of inertia of the nanodiamond. This framework also opens up the possibility of exploring large spatial superpositions induced by controlled rotation along the NV axis, as discussed in~\cite{Rademacher:2025sye}. In addition, this approach can be generalized to systems with more than one NV center, as in~\cite{Braccini:2023eyc}, thereby broadening the scope of applicability of our results.}


\begin{acknowledgments}
We also wish to thank Anom Agung and Qian Xian for many insightful and engaging discussions. R.R. is supported by Beasiswa Indonesia Bangkit and Lembaga Pengelolah Dana Pendidikan (BIB LPDP) of the Ministry of Religious Affairs of Indonesia. R.Z. and T.Z. are supported by the China Scholarship Council (CSC). A.M.'s research is funded by the Gordon and Betty Moore Foundation through Grant GBMF12328, DOI 10.37807/GBMF12328. This material is based on work supported by the Alfred P. Sloan Foundation under Grant No. G-2023-21130.

\end{acknowledgments}

\bibliography{rev1}


\onecolumngrid
\appendix

\section{non-linear Hamiltonian limit}
\label{appendix:non-linear_Hamiltonian}
{
Inverted Harmonic stages are generated by the quadratic magnetic field in Eq. \eqref{eq:non-linear_magnetic_field}, However, in general, the resulting potential will form the quartic curve. we recall Hamiltonian Eq. \eqref{eq:H_general}}
\begin{align*}
    \hat{{H}}_{nl} = \frac{\hat{\textbf{P}}^2}{2m} - \frac{\chi_\rho m}{2\mu_0}\hat{\textbf{B}}^2 +\mu(\hat{\textbf{S}}\cdot\textbf{B})+\hbar D \hat{S}^2_z + \hat{H}_{trap},
\end{align*}
{Here we ignore the term $\mu(\hat{\textbf{S}}\cdot\textbf{B})$ because when the non-linear magnetic field is switched on, the stages that occur are stage enhancement and deceleration when the spin state transitions from $\ket{\pm1}\rightarrow\ket{0}$. Substituting into the non-linear magnetic field Eq. \eqref{eq:non-linear_magnetic_field}, the expression of the potential is simplified to:}
{
\begin{align}
    {{U}}_{nl}(x,y) = \frac{m}{2} (\omega_{x}^2-\omega_{nl}^2)({x}-x_s)^2 + \frac{m}{2} (\omega_{y}^2+\omega_{nl}^2){y}^2 + \frac{m \kappa_{nl}^2}{2} [({x}-x_s)^2+{y}^2]^2 - \frac{\chi_\rho m B_{0(nl)}^2}{2\mu_0}, \label{eq:potential_inverted harmonic}
\end{align}
}
{where $\omega_{nl} = \sqrt{-2\chi_\rho B_{0(nl)}\eta_{nl}/\mu_0}$ and $\kappa_{nl} = \sqrt{-{2\chi_\rho \eta_{nl}^2}/{\mu_0}}$. The presence of the term $\kappa_{nl}$ makes the potential quartic, so we must have the condition $\omega_{nl}^2x^2 \gg \kappa_{nl}^2x^4$ to work in the region of the inverted potential curve. The design proposed by \cite{Elahi:2024dbb} shows trap frequencies of $\omega_x = 2\pi \times 0.01$ Hz and $\omega_y = 2\pi \times 521$ Hz, so it can be assumed that along the $y$-direction the potential remains harmonic. For the superposition direction (the $x$-direction), tuning the magnetic-field bias and gradient at this stage is crucial to maintaining an inverted harmonic curve. In Fig.~\ref{fig:0} (c) shows the potential curve for magnetic-field bias and gradient are set to $B_0 = 0.1$ T and $\eta = 10^6\, \text{T/m}^2$ (based on Ref. \cite{Moorthy:2025bpz}). With this setting, the inverted harmonic region is valid only up to positions of about -200 to 200 $\mu\text{m}$. Therefore, we must choose a such time set for $\tau_2$ so that the quartic term can be neglected, in this case we choose $\tau_2=\tau_4 = 0.25$ s.
}
{The Hamiltonian can be approximated solely for inverted cases as follows
\begin{align}
    \hat{{H}}_{(nl)} = & \ \frac{\hat{{P}}^2_x}{2m} + \frac{\hat{{P}}^2_y}{2m} - \frac{m}{2} (\omega_{nl}^2 -\omega_{x}^2)(\hat{x}-x_s)^2 + \frac{m}{2} (\omega_{nl}^2 + \omega_{y}^2)\hat{y}^2   - \frac{\chi_\rho m B_{0(nl)}^2}{2\mu_0} + \hbar D \hat{S}^2_z.
\end{align}
}

\section{Analytical solutions in every stage}
\label{appendix:analitical_solution}
	
As this scheme consists of five stages, with the following details:

{	
\textbf{STAGE 1} at this stage, a linear magnetic field is activated so that the wave packet will undergo separation into two different spins. The expectation value at this stage uses the solution of equations \eqref{eq:x_linear} and \eqref{eq:x_non-linear} with the initial conditions $\langle \hat{x}(t=0) \rangle = 0$, $\langle \hat{y}(t=0) \rangle = y_0$ \footnote{for non-trivial solution}, $\langle \dot{\hat{x}}(t=0) \rangle = 0$, and $\langle \dot{\hat{y}}(t=0) \rangle = 0$, so the solution to these second-order differential equations can be written as follows (it should be noted, based on Table \ref{tab:magnetic_field_parameters}, 
\begin{align}
	\langle \hat{x} (t) \rangle_1 & = \left(\frac{s \mu \eta_{l}}{m} - \frac{\chi_\rho B_{0(l)} \eta_l}{\mu_0} \right)\frac{[\cos{(\omega_l t) -1}]}{\omega_l^2} \\
	\langle \hat{y} (t) \rangle_1 & = y_0 \cos{\left(\omega_{ly}t\right)}
\end{align}
}
{where $\omega_{ly} = \sqrt{\omega_y^2+\omega_l^2}$ and the value of $t = t_1$ is set as the time when the size of the superposition reaches its maximum condition, 
\begin{align}
	\langle \hat{x}(t_1) \rangle &= x_s-\frac{2s \mu \eta_{l}}{m \omega_l^2} , && \label{eq:stage_1_t1_x}
	\langle \dot{\hat{x}}(t_1) \rangle = 0, \\
	\langle \hat{y}(t_1) \rangle &= y_0 \cos{\left(\omega_{ly} t_1 \right)}, && %
	\langle \dot{\hat{y}}(t_1) \rangle = -y_0\omega_{ly}\sin{\left(\omega_{ly} t_1\right)} \label{eq:stage_1_t1_vy}
\end{align}
}
{Because we want to maintain a symmetric trajectory for both SGI arms in stages 2, 3, and 4, so that the x-path does not cross the boundary of the inverted potential curve, we set the transition time to $t_1 = \pi/\omega_l$. At this time, both trajectories reach a velocity of 0 m/s, and the midpoint of the two arms is exactly at $x_s = -2B_{0(l)}/\eta_l$, which will serve as the symmetry point for stages 2, 3, and 4.}
	
{\textbf{STAGE 2}, the solution for the expectation values at this stage refers to the solution of equations \eqref{eq:x_non-linear} and \eqref{eq:y_non-linear}, which have the initial condition following the maximum superposition time of the first stage in Eq. \eqref{eq:stage_1_t1_x} - \eqref{eq:stage_1_t1_vy}, as  
\begin{align}
	\langle \hat{x}(t) \rangle_2 & = x_s-\frac{2s \mu \eta_{l}}{m \omega_l^2}\cosh{[\omega_{nl} (t-t_1)]} \\
	\langle \hat{y}(t) \rangle_2 & =  \langle\hat{y}(t_1)\rangle
	\cos{\big[\omega_{nly}(t-t_1)\big]}
	+ \frac{\langle\dot{\hat{y}}(t_1)\rangle}{\omega_{nly}}
	\sin{\big[\omega_{nly}(t-t_1)\big]}
\end{align}
}
{where $\omega_{nly} = \sqrt{\omega_y^2+\omega_{nl}^2}$, we can rewrite the y solution in amplitude-phase from
\begin{align}
	\langle \hat{y}(t) \rangle_2 & =  \sqrt{\langle \hat{y}(t_1) \rangle^2 + \left[{\langle \dot{\hat{y}}(t_1) \rangle}/{\omega_{nly}} \right]^2 } \sin{\left[ \omega_{nly} (t-t_1)  -\tilde{\phi}^y_2 \right]}
\end{align}
}
{where $\tilde{\phi}^y_2 = \arcsin\left\{\langle \hat{y}(t_1)/\sqrt{\langle \hat{y}(t_1) \rangle^2 + \left[{\langle \dot{\hat{y}}(t_1) \rangle}/{\omega_{nly}} \right]^2 } \right\}$, at the end of this stage, $t=t_2$, we have
\begin{align}
	\langle \hat{x}(t_2)\rangle &= x_s
	- \frac{2s\mu\eta_l}{m\omega_l^2}\cosh{(\omega_{nl}\tau_2)},\\
	\langle \dot{\hat{x}}(t_2)\rangle &=
	-\frac{2s\mu\eta_l}{m\omega_l^2}\omega_{nl}
	\sinh{(\omega_{nl}\tau_2)},\\
	\langle \hat{y}(t_2)\rangle &= \langle\hat{y}(t_1)\rangle
	\cos{{(\omega_{nly}\tau_2)}}
	+ \frac{\langle\dot{\hat{y}}(t_1)\rangle}{\omega_{nly}}\sin{(\omega_{nly}\tau_2)},\\
	\langle \dot{\hat{y}}(t_2)\rangle &= -\langle\hat{y}(t_1)\rangle\,
	\omega_{nly}\sin{{(\omega_{nly}\tau_2)}}
	+ \langle\dot{\hat{y}}(t_1)\rangle
	\cos{{(\omega_{nly}\tau_2)}}.
\end{align}
where $\tau_2 = t_2 - t_1$.}

{\textbf{STAGE 3}, {we switch it back to the linear magnetic field shifted at the symmetrical point $\mathbf{B}_l(x,y) =\eta_l(x-x_s)\hat{e}_x-\eta_l t\hat{e}_y$ with zero bias (The reason for having a bias in stages 1 and 5 is to fulfill the condition to separate the spin states $\pm1$ and to ensure that the superposition of these two spins is not mixed by the transverse field, a phenomenon known as the Majorana spin-flip. In stage 3, these two requirements are not needed because the spin is in the 0 state, which has no magnetic moment. Moreover, in this stage both arms are never in a zero field region, so it is safe to switch off the bias.)}, make the oscillation around $x_s$:
\begin{align}
	\langle \hat{x}(t)\rangle_3 &= x_s
	+ \big[\langle\hat{x}(t_2)\rangle - x_s\big]\cos\big[\omega_l(t-t_2)\big]
	+ \frac{\langle\dot{\hat{x}}(t_2)\rangle}{\omega_l}
	\sin\big[\omega_l(t-t_2)\big],\\
	\langle \hat{y}(t)\rangle_3 &= \langle\hat{y}(t_2)\rangle
	\cos\big[\omega_{ly}(t-t_2)\big]
	+ \frac{\langle\dot{\hat{y}}(t_2)\rangle}{\omega_{ly}}
	\sin\big[\omega_{ly}(t-t_2)\big].
\end{align}
in the ampitude-phased form 
\begin{align}
	\langle \hat{x}(t) \rangle_3 & = x_s + \sqrt{\big[\langle\hat{x}(t_2)\rangle - x_s\big]^2
	+ \left[\frac{\langle\dot{\hat{x}}(t_2)\rangle}{ \omega_l}\right]^2}\sin{\big[\omega_l(t-t_2) + \tilde{\phi}_3^x\big]}
\end{align}
where  and $\tilde{\phi}_3^x= \arcsin\Bigg\{{[\langle \hat{x}(t_2) \rangle-x_s] /\sqrt{{[\langle \hat{x}(t_2) \rangle-x_s]^2 + \left[{\langle \dot{\hat{x}}(t_2) \rangle}/{\omega_l} \right]^2}}}\Bigg\}$. In this stage, the two trajectory will reach the maximum superposition when $\sin{(\omega_l (t-t_2)+\tilde{\phi}_3^x)} = 1$, so that the maximum superposition only depends on the amplitude of $\langle \hat{x}(t) \rangle_3$, or
\begin{align}
	\Delta x_{max} & = \frac{4\mu \eta_l}{m\omega_{l}^2}\sqrt{\cosh^2{{(\omega_{nl}\tau_2)}}+\left(\frac{\omega_{nl}}{\omega_l} \right)^2 \sinh^2{{(\omega_{nl}\tau_2)}}}
\end{align}
The duration of this stage is fixed by the momentum-reversal condition
$\omega_l\tau_3 = 2 T_0 \mod{(2\pi)}$ with
$T_0 = 2 \arctan{\big[\langle\dot{\hat{x}}(t_2)\rangle/\omega_l,\,
\langle\hat{x}(t_2)\rangle - x_s\big]}$, which yields
\begin{align}
	\langle\hat{x}(t_3)\rangle = \langle\hat{x}(t_2)\rangle,\qquad
	\langle\dot{\hat{x}}(t_3)\rangle = -\langle\dot{\hat{x}}(t_2)\rangle,
\end{align}
The same duration applied to $y$ gives $\langle\hat{y}(t_3)\rangle,
\langle\dot{\hat{y}}(t_3)\rangle$ in the elapsed-time form above with
$t \to t_3$.}

{\textbf{STAGE 4} similar to stage 2, but this time we call it the deceleration stage, we have solutions
\begin{align}
	\langle \hat{x}(t) \rangle_4 & = x_s+ \left[\langle \hat{x}(t_3)\rangle-x_s\right] \cosh{[\omega_{nl}( t- t_3)]}+\frac{\langle \dot{\hat{x}}(t_3) \rangle}{\omega_{nl}}\sinh{[\omega_{nl}( t- t_3)]} \nonumber \\
    & = x_s - \frac{2s\mu\eta_l}{m\omega_l^2}\cosh{[\omega_{nl}(\tau_2-t+t_3)]}\\
	\langle \hat{y}(t) \rangle_4 & = \langle \hat{y}(t_3)\cos{\left[ \omega_{nly} (t - t_3) \right]}+\frac{\langle \dot{\hat{x}}(t_3) \rangle}{\omega_{nl}}\sin{\left[ \omega_{nly} (t - t_3) \right]}
\end{align}
we set the positions and velocities at the end of this stage similar to condition in stage 2, so we choose the time transition $\tau_4 = \tau_2$, which gives
\begin{align}
	\langle\hat{x}(t_4)\rangle = x_s - \frac{2s\mu\eta_l}{m\omega_l^2}, && \langle\dot{\hat{x}}(t_4)\rangle = 0,
\end{align}
the positions of two arms will return to $x_s$ with 0 velocity and prepared to close it in the last harmonic stage.}

{\textbf{STAGE 5} the final stage or recombination stage, which is the same as the first stage (including bias), so that
\begin{align}
	\langle \hat{x}(t) \rangle_5 & =\left( x_s - \frac{2s\mu\eta_l}{m\omega_l^2}\right) \left(1+\cos{[\omega_l( t-t_4)]} \right)\\
	\langle \hat{y} (t) \rangle_5 & =  \langle \hat{y}(t_4) \rangle \cos{\left(\omega_{ly}t-\phi_5^y\right)} +\frac{\langle \dot{\hat{y}}(t_4) \rangle}{\omega_{ly}} \sin{\left(\omega_{ly}t-\phi_5^y\right)}
\end{align}
to close both trajectories, we set $t_5 = t_4+\pi/\omega_l$, resulting in $\langle \hat{x}_R (t_5) \rangle = \langle \hat{x}_L (t_5) \rangle$ and $\langle \dot{\hat{x}}(t=t_5) \rangle = 0$ closing both trajectories at the origin point.}

\section{Euler angle in a spherical nanorotor}
\label{appendix:euler_angle}

\begin{figure}
    \centering
    \includegraphics[width=0.5\linewidth]{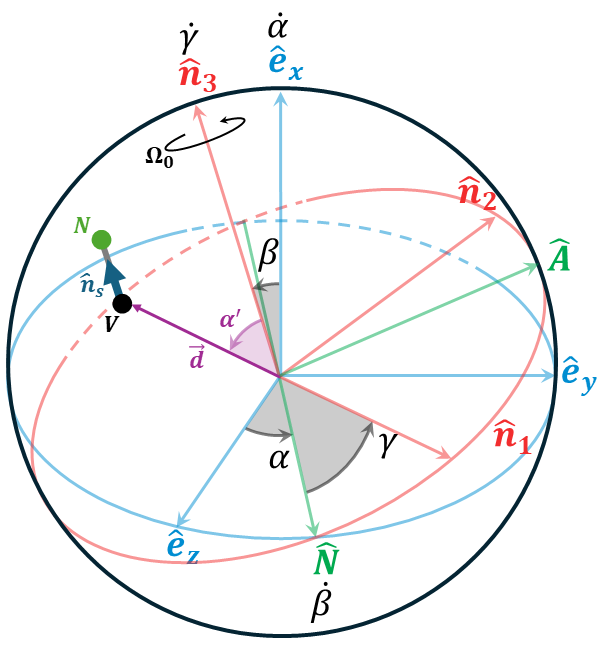}
    \caption{The Euler angles ($\alpha, \beta, \gamma$) of a spherical nanodiamond are represented in the XZX convention, with rotational coordinates ($\hat{n}_1, \hat{n}_2, \hat{n}_3$) and fixed coordinates ($\hat{e}_x, \hat{e}_y, \hat{e}_z$), and $\hat{N}$ denoting the line of nodes. Inside the nanodiamond, there is an embedded NV spin ($\hat{n}_s$) aligned with the $\hat{n}_3$ axis, and its off-center displacement is $\vec{d}$ with a fixed angle $\alpha'$. Following \cite{Zhou:2024pdl}, we assign an initial rotation $\Omega_0$ and an initial libration angle $\beta(t=0) = \beta_0$ to the nanodiamond.}
    \label{fig:euler}
\end{figure}

For the spatial motion of the spherical nanorotor, the applicable Hamiltonian is as follows:
\begin{align}
    \hat{\mathbf{H}}_{s} = \frac{\hat{\mathbf{P}}^2}{2m} + \frac{\hat{\mathbf{L}}^2}{2I} - \frac{\chi_\rho m}{2\mu_0}\mathbf{{B}}^2 + \mu(\hat{\mathbf{S}}\cdot{\mathbf{B}})+D\hat{S}_\parallel^2+\hat{H}_{trap}.
    \label{eq:Hamiltonian_spatial}
\end{align}
In Fig. \ref{fig:euler}, The co-moving frame coordinate of sphere is described in the set vector: $\hat{n}_1$, $\hat{n}_2$, $\hat{n}_3$ and the set vector { $\hat{e}_x, \hat{e}_y, \hat{e}_z$} as a laboratory frame. Therefore, the formulation of angular velocities is given by $\Omega=\dot{\alpha} \hat{e}_x +\dot{\beta} \hat{N}+\dot{\gamma}\hat{n}_3$, where $\hat{N}=\cos{\gamma} \hat{n}_1-\sin{\gamma \hat{n}_2}, \hat{e}_x = \sin{\beta} \hat{A} + \cos{\beta} \hat{n}_3, \hat{A} = \sin{\gamma} \hat{n}_1 +\cos{\gamma} \hat{n}_2$. The angular velocity at the co-moving frame can be written as
\begin{align}
    \Omega_1 = \dot{\alpha} \sin{\beta} \sin{\gamma} + \dot{\beta} \cos{\gamma}, &&
    \Omega_2 = \dot{\alpha} \sin{\beta} \cos{\gamma} - \dot{\beta} \sin{\gamma}, &&
    \Omega_3 = \dot{\alpha} \cos{\beta} + \dot{\gamma}.
\end{align}
The rotational kinetic energy of the spherical nanorotor ($T_{rot} = \frac{1}{2}\Sigma_{i=1}^3 I_i \Omega^2_i $) can be formulated as
\begin{align}
    T_{rot} = \frac{I}{2}\left[ \dot{\alpha}^2\sin^2\beta +\dot{\beta}^2+(\dot{\alpha}\cos{\beta} + \dot{\gamma})^2\right] \label{eq:kinetic_rotation}
\end{align}
we can also write the kinetic energy in the canonical momentum form ($p_q = \partial T/\partial\dot{q}$) as
\begin{align}
    p_\alpha &= I(\dot{\alpha}+\dot{\gamma}\cos\beta) \label{eq:momenta_alpha} \\
    p_\beta &=I\dot{\beta} \label{eq:momenta_beta}\\ 
    p_\gamma &= I(\dot{\gamma}+\dot{\alpha}\cos\beta) \label{eq:momenta_gamma}
\end{align}
because there is no external torque along $\alpha$ and $\gamma$, except for the Zeeman effect, which depends on $\beta$. Therefore, $\dot{p}_\alpha =\dot{p}_\gamma = 0$, so that $p_\alpha=L_x$ and $p_\gamma = L_3$ are constant. We can get a Hamiltonian form in Eq. \eqref{spin-Ham0} using Eq. \eqref{eq:kinetic_rotation} - \eqref{eq:momenta_gamma}.

The equations of motion in the $x$ and $y$ directions, as derived from the Hamiltonian in Eq. \eqref{eq:Hamiltonian_spatial}, for the linear part, are
\begin{align}
    \frac{d^2\langle{x}\rangle_l}{dt^2} & = \frac{\chi_\rho}{2 \mu_0} \frac{\partial \mathbf{B}^2}{\partial x} - \frac{\mu}{m} S_\parallel \frac{\partial B_\parallel}{\partial x} \label{eq:x_general}\\
    \frac{d^2\langle{y}\rangle_l}{dt^2} & = \frac{\chi_\rho}{2 \mu_0} \frac{\partial \mathbf{B}^2}{\partial y} - \frac{\mu}{m} S_\parallel \frac{\partial B_\parallel}{\partial y} \label{eq:y_general},
\end{align}
Given that the NV spin is aligned along $\hat{n}_3$, whose projection on the fixed coordinate system is given by
$\hat{n}_s = \{\cos{\beta}, \sin{\beta},0\}$, and the magnetic field profile applied in this setup can generally be expressed as $\mathbf{B} = \{B_x, B_y, 0\}$, the component of the magnetic field along the NV axis, $\hat{n}_s$, is the projection $B_\parallel = \textbf{B} \cdot \hat{n}_s$.
Under these conditions, the equation of motion can be formulated as
\begin{align}
    \frac{d^2\langle{x}\rangle_l}{dt^2} & = \frac{\chi_\rho}{2 \mu_0} \frac{\partial \mathbf{B}^2}{\partial x} - \frac{\mu s}{m}  \left(\cos{\beta} \frac{\partial B_x}{\partial x}\right) \label{eq:z}\\
    \frac{d^2\langle{y}\rangle_l}{dt^2} & = \frac{\chi_\rho}{2 \mu_0} \frac{\partial \mathbf{B}^2}{\partial y} - \frac{\mu s}{m} \left( \sin{\beta} \frac{\partial B_y}{\partial y} \right), \label{eq:y}
\end{align}
where $s = \{-1,+1\}$ is considered the spin state in the stages of the linear magnetic field.

\section{Wave packet evolution}
\label{appendix:WPE_general}
\subsection{Spatial}
\label{appendix:Spatial Contrast}

First, we formulate the evolution of the wave packet. In this case, we consider both the harmonic potential and the inverted harmonic potential. This formalism follows \cite{Zhou:2024voj}; however, we include the bias magnetic field $B_{0(l)}$ in the general trajectory solution, particularly during the harmonic potential stage, so that the Lagrangian formation becomes:
\begin{align}
	\mathcal{L} (t)= \frac{1}{2} m\dot{x}^2(t) -  \frac{1}{2} m\omega_l^2 x^2(t) +A_0 x(t)
\end{align}
where $A_0 = S_x \mu \eta_l-\chi_\rho m B_{0(l)} \eta_{l}/\mu_0$, we have the general trajectory solution for the harmonic potential as
\begin{align}
	x(t) = \frac{A_0}{m\omega_{l}^2}+\left(x_0-\frac{A_0}{m\omega_{l}^2}\right)\cos{(\omega_l t)}+\frac{p_0}{m\omega_{l}}\sin{\omega_l t}
\end{align}
where $x_0$ and $p_0$ are the initial position and momentum. If we have the boundary conditions $x(0) =x_i$ and $x(t_f) = x_f$, we have
\begin{align}
	x(t) = \frac{A_0}{m\omega_{l}^2}+\left(x_i-\frac{A_0}{m\omega_{l}^2}\right)\cos{(\omega_l t)}+\frac{\left(x_f - \frac{A_0}{m\omega_{l}^2}\right)-\left(x_i-\frac{A_0}{m\omega_{l}^2}\right)\cos{(\omega_l t_f)}}{\sin{(\omega_l t_f)}}\sin{(\omega_l t)}
\end{align}
Now we can formulate the action as follows
\begin{align}
	S = & \int_{t_i}^{t_f} dt \left\{ \frac{1}{2} m\dot{x}^2(t) - \frac{1}{2} m\omega_l^2 x^2(t) + A_0 x(t) \right\}, \\
	=& \frac{m\omega_l}{2 \sin(\omega_l t)}\left[(x^2_f + x^2_i)\cos(\omega_l t) - 2x_f x_i\right]+\frac{A_0(x_f + x_i)}{\omega_l \sin(\omega t)} \left[1-\cos(\omega_{l} t)\right] + \frac{A_0^2}{2m\omega_l^2}\left\{t - \frac{[1-\cos(\omega_l t)]}{\omega_{l}\sin{(\omega_l t)}}\right\}
\end{align}
We used path integral to formulate the evolution of wave function,
\begin{align}
	\psi(x,t)=\int dx' K(x,t;x',0)\psi(x',0), \label{eq:gaussian formulation}
\end{align}
where $\psi(x',0)$ is the initial wave function and $K(x,t;x',0)$ known as the propagator can be solved using the Van Vleck-Pauli-Morette formula:
\begin{align}
	K(x_f,t_f;x_i,t_i) = \sqrt{\frac{i}{2\pi \hbar} \frac{\partial^2 S}{\partial x_f \partial x_i}}\exp[\frac{iS}{\hbar}], \label{eq:VPM formula}
\end{align}
in our case, we get
\begin{align}
	K(x,t;x',0) =& \sqrt{\frac{m \omega_l}{2\pi i\hbar \sin(\omega_l t)}}\exp\Bigg\{\frac{i}{\hbar}  \frac{m\omega_l}{2 \sin(\omega_l t)}\left[(x^2 + x'^2)\cos(\omega_l t) - 2x x'\right] \nonumber \\
	&+\frac{i}{\hbar}\frac{A_0(x_f + x_i)}{\omega_l \sin(\omega t)} \left[1-\cos(\omega_{l} t)\right] +\frac{i}{\hbar} \frac{A_0^2}{2m\omega_l^2}\left[t - \frac{(1-\cos(\omega_l t))}{\omega_{l}\sin{(\omega_l t)}}\right] \Bigg\}, \label{eq:initial gaussian function}
\end{align}
because we used a quadratic form of the Gaussian wave packet, so the initial wave packet became
\begin{align}
	\psi(x',0) = N_0 \exp\left[-\frac{(x'-x_0)^2}{4\sigma_0^2}+i\left(\frac{a_0}{4} x'^2 + b_0 x' + c_0\right)\right]
\end{align}
combined Eqs. \eqref{eq:initial gaussian function}, \eqref{eq:VPM formula}, and \eqref{eq:gaussian formulation}, we have
\begin{align}
	\psi(x,t) = N(t) \exp\left[i\frac{x^2}{4u^2(t)}-\frac{x_0^2}{4\sigma^2_0}+ \frac{A_0(1-2\cos(\omega_l t))}{\omega_l \sin(\omega_l t)}x\right] \exp\left[\frac{\left(ib_0-i\frac{x}{2u^2(t)\cos(\omega_l t)}+ \frac{x_0}{2\sigma_0^2} +\frac{A_0}{\omega\sin(\omega_l t)} \right)^2}{1/\sigma_0^2 - i(a_0+1/u^2(t))}\right]
    \label{eq:wave_packet_linear_psi}
\end{align}
where
\begin{align}
	u(t) & = \sqrt\frac{\hbar \sin(\omega_l t)}{2m\omega_l\cos(\omega_l t)}, \\
	N(t) & = N_0 \sqrt{\frac{m \omega_l}{2\pi i\hbar \sin(\omega_l t)}} \sqrt{\frac{4\pi}{1/\sigma_0^2 -i(a_0+1/u^2(t))}} \exp{i\left(\frac{A_0^2}{2m\hbar\omega_l^2}\left[t - \frac{(1-\cos(\omega_l t))}{\omega_{l}\sin{(\omega_l t)}}\right]+c_0\right)}
\end{align}
The wave packet equation \eqref{eq:wave_packet_linear_psi} can be written in the following GSWP form.
\begin{align}
	\psi(x,t) = N(t) \exp[-\frac{1}{4\sigma^2(t)}(x-x_c(t))^2+i\left(\frac{a(t)}{4}x^2+b(t)x+c(t)\right)]
\end{align}
where
\begin{align}
	\sigma(t) =& \sigma_0\left[\frac{\hbar^2}{4m^2\omega^2_l \sigma_0^2} \sin^2(\omega_l t) +\left(\frac{a_0 \hbar}{2m \omega_l} \sin(\omega_l t) + \cos(\omega_l t)\right)^2 \right]^{\frac{1}{2}} \\
    x_c(t)= & \frac{\hbar a_0 x_0}{2m\omega_i} \sin{(\omega_it)}+x_0 \cos{(\omega_l t)}+\frac{\hbar b_0}{m\omega_l} \sin{(\omega_l t)}+\frac{A_0(1-\cos{(\omega_l t)})}{m \omega_l^2}
\end{align}
In the GSWP equation, the value $\sigma(t)$ represents the evolution of the wave packet width, and $x_c$ is the classical equation of motion of the nanodiamond along the dimension x. If we set $A_0 = 0$, the $x_c$ reduces to Eq. (63) in ref. \cite{Zhou:2024voj}. It can be concluded that adding a magnetic field bias term to the magnetic field profile in the linear stage affects the classical equation of motion of the nanodiamond but does not affect the width of its wave packet peak. Finally, for the imaginary parameters, define as:
\begin{align}
    a(t) &= \frac{1}{u_t^2} - \frac{1 + a_0 u_t^2}{4 u_t^6 \cos^2(\omega_l t) \left[ \left( \frac{1}{u_t^2} + a_0 \right)^2 + \frac{1}{\sigma_0^4} \right]} \\
    b(t) &= \frac{2 b_0 \sigma_0^4 - u_t^2 \left( x_0 - 2 a_0 b_0 \sigma_0^4 \right)}{2 \cos(\omega_l t) \left[ \sigma_0^4 + 2 a_0 u_t^2 \sigma_0^4 + u_t^4 \left(1 + a_0^2 \sigma_0^4 \right) \right]} + b_A(A_0,t) \\
    c(t) &= \frac{x_0^2 + x_0 u_t^2 \left(4 b_0 + a_0 x_0 \right) - 4 b_0^2 \sigma_0^4 \left(1 + a_0 u_t^2 \right)}{4 u_t^2 \sigma_0^4 \left[ \left( \frac{1}{u_t^2} + a_0 \right)^2 + \frac{1}{\sigma_0^4} \right]} + c_A(A_0,t)
\end{align}
where
\begin{align}
    b_A(A_0,t) = & \frac{A_0\left[ 2 m {\sigma_0}^4 u^2 \omega_l  \left({a_0} u_t^2+1\right) \csc (\omega_l t  )+\hbar  \left({\sigma_0}^4 \left({a_0} u^2+1\right)^2+u_t^4\right) \cos (\omega_l t )-\hbar  \left({\sigma_0}^4 \left({a_0} u_t^2+1\right) \left({a_0} u_t^2+2\right)+u_t^4\right) \right]}{\omega_l \hbar \sin(\omega_l t) \left[ \sigma_0^4 + 2 a_0 u_t^2 \sigma_0^4 + u_t^4 \left(1 + a_0^2 \sigma_0^4 \right) \right]} \\
    c_A(A_0,t) = & -\frac{{A_0} [\sec (\omega_l t)-1] \left[{u_t}^2 {x_0}-2 {b_0} {\sigma_0}^4 \left({a_0} {u_t}^2+1\right)\right]}{2 m \omega_l^2 u_t^4 \sigma_0^4 \left[ \left( \frac{1}{u_t^2} + a_0 \right)^2 + \frac{1}{\sigma_0^4} \right]} +O(A_0^2).
\end{align}
Next, the solution for the IHP is derived in the same way, but with Lagrangian
\begin{align}
	\mathcal{L} (t)= \frac{1}{2} m\dot{x}^2(t) + \frac{1}{2} m\omega_{nl}^2 x^2(t)
\end{align}
Thus, an identical solution is obtained, but in terms of hyperbolic functions. The solution is obtained as follows:
\begin{align}
	\sigma^I(t) =& \sigma_0\left[\frac{\hbar^2}{4m^2\omega^2_l \sigma_0^2} \sinh^2(\omega_{nl} t) +\left(\frac{a_0 \hbar}{2m \omega_l} \sinh(\omega_{nl} t) +  \cosh(\omega_{nl} t)\right)^2 \right]^{\frac{1}{2}} \\
    x_c^I(t)= & \frac{\hbar a_0 x_0}{2m\omega_i} \sinh{(\omega_{nl}t)}+x_0 \cosh{(\omega_{nl}t)}+\frac{\hbar b_0}{m\omega_{nl}} \sinh{(\omega_{nl}t)}
\end{align}
and for imaginary part coefficient, we have:
\begin{align}
    a^I(t) &= \frac{1}{{u_t^I}^2} - \frac{1 + a_0 {u_t^I}^2}{4 {u_t^I}^6 \cosh^2(\omega_{nl} t) \left[ \left( \frac{1}{{u_t^I}^2} + a_0 \right)^2 + \frac{1}{\sigma_0^4} \right]} \\
    b^I(t) &= \frac{2 b_0 \sigma_0^4 - {u_t^I}^2 \left( x_0 - 2 a_0 b_0 \sigma_0^4 \right)}{2 \cosh(\omega_{nl} t) \left[ \sigma_0^4 + 2 a_0 {u_t^I}^2 \sigma_0^4 + {u_t^I}^4 \left(1 + a_0^2 \sigma_0^4 \right) \right]} \\
    c^I(t) &= \frac{x_0^2 + x_0 {u_t^{I}}^2 \left(4 b_0 + a_0 x_0 \right) - 4 b_0^2 \sigma_0^4 \left(1 + a_0 {u_t^I}^2 \right)}{4 {u_t^I}^2 \sigma_0^4 \left[ \left( \frac{1}{{u_t^I}^2} + a_0 \right)^2 + \frac{1}{\sigma_0^4} \right]}
\end{align}
where
\begin{align}
	{u^I}(t) & = \sqrt \frac{\hbar \sinh(\omega_{nl} t)}{2m\omega_{nl}\cosh(\omega_{nl} t)}, \\
    N(t) & = N_0 \sqrt{\frac{m \omega_{nl}}{2\pi i\hbar \sinh(\omega_{nl} t)}} \sqrt{\frac{4\pi}{1/\sigma_0^2 -i(a_0+1/{u^I}^{2}(t))}} e^{ic_0}
\end{align}

\subsection{rotation}
Because the solution of the libration mode is always a harmonic oscillation (which is $\beta$ has a minimum value to be gyroscopically stable), one can assume that the wave packet will be of Gaussian form as in Ref. \cite{Zhou:2024pdl}. Before calculating the wave packet, we first apply a shift to the operators by translating the angular and momentum variables around their expectation values, thereby simplifying the analysis of quantum fluctuations and focusing on small deviations from the equilibrium position. By treating the mean values as the classical component of the quantum fluctuations, the dynamical modeling becomes simpler since the system can be regarded as a small harmonic oscillator around its equilibrium point. This approach also reduces the complexity of the Hamiltonian by eliminating linear terms, redirecting attention to the variance and coherence of the quantum wave. Moreover, it enables for a more accurate calculation of the overlap between the quantum waves in the two interferometer paths. The shifts are defined as follows {\footnote{Here, the canonical quantization procedure (see Ref. \cite{Barut1992}). We have the commutation relations of the angular momentum: $[\hat{\beta},\hat{p}_\beta]= [\hat{\alpha},\hat{p}_\alpha] = [\hat{\gamma},\hat{p}_\gamma] = i\hbar \ \text{and} \ [\hat{p}_\beta,\hat{p}_\alpha] = [\hat{p}_\beta,\hat{p}_\gamma] = [\hat{p}_\alpha,\hat{p}_\gamma] = 0$}}:
\begin{align}
	\hat{\beta}' &\equiv \hat{\beta}-\bar{\beta}, \\
	\hat{p}'_q &\equiv \hat{p}_q - \langle \hat{p}_q \rangle, q = \{\alpha,\gamma\}
\end{align}
The expectation values of the momentum operators for the two angular coordinates are defined as $\langle \hat{p}_\alpha \rangle = I\Omega_0\cos{\beta_0}$ dan $\langle \hat{p}_\gamma \rangle = I\Omega_0$. Recall the rotational Hamiltonian:
\begin{align}
	\hat{H}_{rot} & 
    \approx \frac{\hat{p}_\beta^2}{2I}+\frac{I\Omega_0^2}{2}\left[{\beta'-\frac{f(\hat{p}'_\alpha,\hat{p}'_\gamma)}{I\Omega_0^2}}\right]^2 - \frac{f^2(\hat{p}'_\alpha,\hat{p}'_\gamma)}{2I\Omega_0^2}+g(\hat{p}'_\alpha,\hat{p}'_\gamma,t) \label{eq:Hamiltonian_rotation_approx},
\end{align}
where
\begin{align}
	f(\hat{p}'_\alpha,\hat{p}'_\gamma) & =  \frac{(\hat{p}'_\alpha-\hat{p}'_\gamma )^2}{I{\beta_0^3}} -\frac{\Omega_0}{\beta_0}{(\hat{p}'_\alpha-\hat{p}'_\gamma )} , \\
	g(\hat{p}'_\alpha,\hat{p}'_\gamma,t) & = \frac{(\hat{p}'_\alpha-\hat{p}'_\gamma )^2}{2I{\beta_0^2}}  + \frac{\mu s}{I \Omega_0 \beta_0}[B_x(t)\beta_0-B_y(t) + d \eta_l \sin{\alpha_0}]{(\hat{p}'_\alpha-\hat{p}'_\gamma )} + \Omega_0\hat{p}'_\gamma,
\end{align}
we can neglect the high-order term of $\beta'$ based on the assumption $\beta' \ll \beta_0 \ll 1$. Here we introduce $\mathcal{B}(t) = B_x(t) \beta_0 - B_y(t) + d \eta_l \sin \alpha_0$, which was neglected for all terms except, $B_x(t) \beta_0$, in Refs. \cite{Zhou:2024pdl, Rizaldy2025_RotationalStability} because the distance of NV off-center of the NV ($d \eta_l \sin \alpha_0$) is not significant enough. However, since we apply a high-gradient magnetic field, we must include this term. 

First, when the interferometer has not yet been activated, the spin state of the NV nanodiamond is in the state $s = 0$. At this moment, the quantum state of the libration mode is in the ground state $\ket{0}_{\bar{\beta}_0}$, which lies at the trapping frequency $\Omega_0$ and the equilibrium position $\bar{\beta}_0$. Consequently, the wave packet in this state is purely the wave packet of the precession mode $\alpha$ and the rotation mode $\gamma$, with wave-packet widths $\sigma_{p_\alpha}$ and $\sigma_{p_\gamma}$, as follow
\begin{align}
    \ket{\Psi(0)} = & \ \int \frac{dp'_\alpha dp'_\gamma}{\sqrt{2\pi \sigma_{p_\alpha} \sigma_{p_\gamma}}} e^{-\frac{p'^2_\alpha}{4\sigma_{p_\alpha}}} e^{-\frac{p'^2_\gamma}{4\sigma_{p_\gamma}}} \ket{p'_\alpha,p'_\gamma} \otimes \ket{0}_{\bar{\beta}_0}.
\end{align}
According to the Hamiltonian \eqref{eq:Hamiltonian_rotation_approx},  the equilibrium of $\beta$ in the ground state is defined as 
\begin{align}
    \bar{\beta}_0(p'_\alpha,p'_\gamma) = \beta_0 + \frac{f(p'_\alpha,p'_\gamma)}{I\omega_0^2}
    \label{eq:initial_equilibrium_position}.
\end{align}
The additional term \({f(p'_\alpha, p'_\gamma)}/{I\Omega_0^2}\) represents a position shift due to fluctuations in the \(\alpha\) and \(\gamma\) momenta. Immediately after the interferometer is activated, the spin state of the NV nanodiamond is in superposition, changing from \(s = 0\) to \(s = +1\) and \(s = -1\) (in stage 1). In this situation, the equilibrium position \(\beta\) is shifted and becomes the following:
\begin{align}
    \bar{\beta}(s,t,p'_\alpha,p'_\gamma) = \beta_0 + \frac{s\mu B(t)\beta_0}{I\Omega_0^2}+\frac{f(p'_\alpha,p'_\gamma)}{I\Omega_0^2}.
    \label{eq:new_equilibrium_position}
\end{align}
Through this change in spin state, we can now formulate the evolution of the wave packet into
\begin{align}
    \ket{\Psi(t)} = & \ \exp{-\frac{i}{\hbar} \int \hat{H}_{rot}} \ket{\Psi(0)} \nonumber \\
    = & \ \int \exp \left[ -\frac{i}{\hbar} \int dt \left( \frac{\hat{p}_\beta^2}{2I}+\frac{I\Omega_0^2}{2}\left[{\beta'-\frac{f(\hat{p}'_\alpha,\hat{p}'_\gamma)}{I\Omega_0^2}}\right]^2 - \frac{f^2(\hat{p}'_\alpha,\hat{p}'_\gamma)} {2I\Omega_0^2}+g(\hat{p}'_\alpha,\hat{p}'_\gamma,t)\right) \right] \ket{\Psi(0)} \nonumber \\
    = &  \ \int \frac{dp'_\alpha dp'_\gamma}{\sqrt{2\pi \sigma_{p_\alpha} \sigma_{p_\gamma}}} e^{-\frac{p'^2_\alpha}{4\sigma_{p_\alpha}^2}} e^{-\frac{p'^2_\gamma}{4\sigma_{p_\gamma}^2}} \times \exp \left[ -\frac{i}{\hbar} \int dt g(p'_\alpha,p'_\gamma,t) \right] \ket{p'_\alpha,p'_\gamma} \otimes \ket{\kappa(t)}_{\bar{\beta}}.
    \label{eq:wave_packet_of quantum_evolution1}
\end{align}
where $\ket{\kappa(t)}_{\bar{\beta}}$ is the coherent state defined as
\begin{align}
    \ket{\kappa(t)}_{\bar{\beta}} &= \int \exp \left[ -\frac{i}{\hbar} \int dt \left( \frac{\hat{p}_\beta^2}{2I}+\frac{I\Omega_0^2}{2}\left[{\beta'-\frac{f(\hat{p}'_\alpha,\hat{p}'_\gamma)}{I\Omega_0^2}}\right]^2 - \frac{f^2(\hat{p}'_\alpha,\hat{p}'_\gamma)}{2I\Omega_0^2}\right) \right] \nonumber \\
    &= \ket{- \sqrt{\frac{I \omega_0}{2\hbar}} \frac{s\mu B(t) \beta_0}{I\omega_0^2}e^{-i\Omega_0 t}}_{\bar{\beta}}.
    \label{eq:coherent_state_def_1}
\end{align}
At recombination time, the coherent state, $\kappa(t_5)$, can be estimated as $\abs{\kappa(t_5)}<\sqrt{I\Omega_0/(2\hbar)} 3\mu B_0 \beta_0/(I\Omega_0^2)$, (See \cite{Zhou:2024pdl} for the detailed derivation). However, the quantum states of the angles $\alpha$ and $\gamma$ maintain their Gaussian form throughout the time evolution (considered the Hamiltonian for that part consisting only of $p'^2_q, q=\{\alpha,\gamma\}$), and since $[p_q,H]=0$, the width of the wave packet of $\sigma_{p_q}$ are fixed.

\section{Humpty-Dumpty Effect at Recombination Stage}
\label{appendix:humpty-dumpty}
\subsection{Spatial}
After we successfully constructed the wave packets function, we now can determine the contrast of these two arms SGI, which are $\psi_L(x,t)$ and $\psi_L(x,t)$ (See Eq. (18)). Because all parameters in Eq. (18) remain the same except the term of classical position, $x_c$, and the classical momentum related parameter, $b(t)$, so that
\begin{align}
	\psi_{\text{L}}(x,t) = N(t)\exp\left[-\frac{(x-x_{\text{L}}(t))^{2}}{4\sigma_{x}^{2}(t)}+ib_{\text{L}}x\right], & & 
	\psi_{\text{R}}(x,t) = N(t)\exp\left[-\frac{(x-x_{\text{R}}(t))^{2}}{4\sigma_{x}^{2}(t)}+ib_{\text{R}}x\right].
\end{align}
The contrast is defined as follows
\begin{align}
	C(t) &= \int_{-\infty}^{\infty} dx\,\psi_{\text{L}}^{*}(x,t)\psi_{\text{R}}(x,t) \nonumber\\
	&= N^{2}(t)\int_{-\infty}^{\infty} dx\,\exp\Bigg[-\frac{1}{4\sigma_{x}^{2}(t)}\bigg(2x^{2} -2\left(x_{\text{R}}(t)+x_{\text{L}}(t)+2i\sigma_{x}^{2}(t)(b_{\text{R}}-b_{\text{L}})\right)x +x_{\text{L}}^{2}(t)+x_{\text{R}}^{2}(t)\bigg)\Bigg].
\end{align}
Using the Gaussian integral formula:
\begin{align}\label{gaussian_int_formula}
	\int_{-\infty}^{\infty} e^{-(ax^2 + bx + c)} \, dx = \sqrt{\frac{\pi}{a}} \, e^{\frac{b^2}{4a} - c},
\end{align}
with the identifications:
\begin{align}
	a(t) = \frac{1}{2\sigma_{x}^{2}(t)}, & &
	b(t) = -\frac{1}{2\sigma_{x}^{2}(t)}\left[x_{\text{R}}(t)+x_{\text{L}}(t)+2i\sigma_{x}^{2}(t)(b_{\text{R}}-b_{\text{L}})\right], & &
	c(t) = \frac{x_{\text{R}}^{2}(t)+x_{\text{L}}^{2}(t)}{4\sigma_{x}^{2}(t)}.
\end{align}
Computing the exponent:
\begin{align}\label{int_mid_step}
	\frac{b^2}{4a}-c
	&= i\phi - \frac{(x_{\text{R}}(t)-x_{\text{L}}(t))^2}{8\sigma_{x}^{2}(t)} - \frac{\sigma_{x}^{2}(t)\Delta b^{2}}{2},
\end{align}
where we have introduced:
\begin{align}
	\phi = \frac{(b_{\text{R}}-b_{\text{L}})(x_{\text{R}}(t)+x_{\text{L}}(t))}{2}, & & 
	\Delta x = x_{\text{R}}(t)-x_{\text{L}}(t), & &
	\Delta b = b_{\text{R}}(t)-b_{\text{L}}(t).
\end{align}
Substituting Eq.~\eqref{int_mid_step} into Eq.~\eqref{gaussian_int_formula}, we obtain the following
\begin{align}
	C(t) = N^{2}(t)\sqrt{2\pi\sigma_{x}^{2}(t)}\exp\left[i\phi - \frac{\Delta x^2}{8\sigma_{x}^{2}(t)} - \frac{\sigma_{x}^{2}(t)\Delta b^{2}}{2}\right].
\end{align}
Ignoring the normalization coefficient and phase factor, the contrast is as follows:
\begin{align}\label{contrast}
	C(t) \approx \exp\left[-\frac{\Delta x^2}{8\sigma_{x}^{2}(t)} - \frac{\sigma_{x}^{2}(t)\Delta b^{2}}{2}\right].
\end{align}

\subsection{Rotation}

We calculated the SGI contrast from the total integral of the overlapping wave function of two-arm interferometry when recombined at $t_5$
\begin{align}
	C = \abs{\braket{\Psi_L(t_5)}{\Psi_R(t_5)}},
\end{align}
where the total wave function is defined as a contribution of every mode $\ket{\Psi} = \ket{\psi_{trans}} \otimes \ket{\psi_{rot}}$. 
If we assume that there is perfect recombination in translation overlap $\braket{\psi_{trans}^L}{\psi_{trans}^R} = 1$ (given the size of the superposition ($\Delta x$) and the velocity difference are $\approx$ 0). Thus, the main limiting factor in our scheme focuses on rotation:
\begin{align}
    C=\braket{\psi_{rot}^L}{\psi_{rot}^R} = \braket{\alpha_L}{\alpha_R}\braket{\beta_L}{\beta_R}\braket{\gamma_L}{\gamma_R} \label{eq:contrast_formalism},
\end{align}
The first and third terms represent the Gaussian wave packet in precession and rotation modes, respectively, and can be formulated as
\begin{align}
    \ket{\alpha} = \int \frac{dp'_\alpha}{\sqrt{2\pi\sigma_{p_\alpha}}} \exp\left[-\frac{(p'_\alpha)^2}{4\sigma_{p_\alpha}^2} \right] \ket{p'_\alpha}, & &
    \ket{\gamma} = \int \frac{dp'_\gamma}{\sqrt{2\pi\sigma_{p_\gamma}}} \exp\left[-\frac{(p'_\gamma)^2}{4\sigma_{p_\gamma}^2}\right] \ket{p'_\gamma} \label{eq:ket_alpha_gamma},
\end{align}
Here, the mismatch of the precession and rotation angle is defined as $\delta \alpha = \alpha_L(t_5) - \alpha_R(t_5)$ and $\delta \gamma = \gamma_L(t_5) - \gamma_R(t_5)$, and $\sigma_{p_\alpha}$ and $\sigma_{p_\gamma}$ represent the uncertainty of the initial angle momentum or the width of the Gaussian wave packet, with the assumption that the quantum fluctuation of the momentum is small, so that $\sigma_{p_\alpha} \approx \sigma_{p_\gamma} \sim \hbar$. Lastly, the wave packet of the librational mode at recombination time is
\begin{align}
  \ket{\beta(t_5)} &= \exp \left[ -\frac{i}{\hbar} \int_0^{t_5} dt g(p'_\alpha,p'_\gamma,t) \right] \ket{p'_\alpha,p'_\gamma} \otimes \ket{\kappa(t)}_{\bar{\beta}}, \label{eq:ket_beta} 
\end{align}
By inserting Eq. \eqref{eq:ket_alpha_gamma} and \eqref{eq:ket_beta} into Eq. \eqref{eq:contrast_formalism}, along with Eq. \eqref{eq:coherent_state_def_1} and following the formalism of Ref. \cite{Zhou:2024pdl}, we obtain:
\begin{align}
    C > \exp\left(-\frac{\delta \alpha^2 \sigma_{p_\alpha}^2}{2 \hbar^2}-\frac{\delta \gamma^2 \sigma_{p_\gamma}^2}{2 \hbar^2}- \frac{16 \mu^2 \mathcal{B}^2}{\hbar I \Omega_0^3}\right)
    \label{eq:contrast_non_T}
\end{align}
where $\mathcal{B}(t_5)\approx(B_{0}^{(l)}\beta_0-y_0 \eta_l+d\eta_l\sin{\alpha_0})$. For spin contrast in the finite temperature case can be considered as $\rho_{th} = \int d^2 \kappa P(\kappa) \ket{\kappa}\bra{\kappa}$, where $P(\kappa)$ represents the distribution function of the thermal state. So, the spin contrast for finite temperature is given by \cite{Zhou:2024pdl}
\begin{align}
    C_{t h}>\exp \left[-\frac{\delta \alpha^2 \sigma_{p_\alpha}^2}{2 \hbar^2}-\frac{\delta \gamma^2 \sigma_{p_\gamma}^2}{2 \hbar^2}- \frac{16(1+2 n) \mu^2 \mathcal{B}^2}{\hbar I \Omega_0^3}\right]. \label{eq:contrast_temprature_appendix}
\end{align}
where $n=k_B T/(\hbar \omega_0)$ defines as the occupation number.

\end{document}